\documentclass{aa}  

\usepackage{graphicx}
\usepackage{txfonts}
\usepackage{subfig}	
\usepackage{amsmath,bm,hyperref,booktabs,multirow}
\usepackage{multirow}
\usepackage{natbib}
\bibpunct{(}{)}{;}{a}{}{,} 
\usepackage{comment}
\usepackage{tablefootnote}
\usepackage{rotating}
\usepackage[dvipsnames]{xcolor}
\hypersetup{colorlinks=true,citecolor=blue,
linkcolor=blue,filecolor=magenta,urlcolor=blue}

\begin{document}

  \title{Str\"{o}mgren photometric metallicity map of the Magellanic Cloud stars using Gaia DR3--XP spectra}
	\titlerunning{Metallicity maps of the Clouds}

   \author{Abinaya O. Omkumar
          \inst{1,2,3}
          \thanks{aomkumar@aip.de}
          \and
          Maria-Rosa L. Cioni\inst{1}  
          \and
          Smitha Subramanian\inst{2,1}
          \and
          Jos de Bruijne\inst{4}
          \and
          Aparna Nair\inst{2,5}
          \and
          Bruno Dias\inst{6}          
          }
          
	\authorrunning{Omkumar et al.}
	
   \institute{Leibniz-Institut f\"ur Astrophysik Potsdam, An der Sternwarte 16, D-14482 Potsdam, Germany 
         \and
          Indian Institute of Astrophysics, Koramangala II Block, Bangalore-560034, India\               
         \and 
            Institut f\"{u}r Physik und Astronomie, Universit\"{a}t Potsdam, Haus 28, Karl-Liebknecht-Str. 24/25, D-14476 Potsdam, Germany
        \and
            ESA, European Space Research and Technology Centre, Keplerlaan 1, 2201 AZ Noordwijk, The Netherlands
           \and 
           Instituto de Astronomia y Ciencias Planetarias , Univeridad de Atacama, Copayapu 485 Copiapo, Chile
           \and
           Instituto de Astrofísica, Departamento de Física y Astronomía, Facultad de Ciencias Exactas, Universidad Andres Bello, Fernandez Concha, 700, Las Condes, Santiago, Chile
        }

   \date{Received ; accepted }

% \abstract{}{}{}{}{} 
% 5 {} token are mandatory
 
  \abstract
  {One important key in understanding a galaxy's evolution is to study the consequences of its past dynamical interactions that influenced its shape. By measuring the metallicity distribution of stellar populations with different ages, one can learn about these interactions. The Magellanic Clouds, being the nearest pair of interacting dwarf galaxies with a morphology characterised by different tidal and kinematic sub-structures as well as a vast range of stellar populations, represent an excellent place to study  the consequences of dwarf-dwarf galaxy interactions and with their large host, the Milky Way.} 
  {We aim to determine the metallicities ([Fe/H]) of red giant branch (old) and supergiant (young) stars covering the entire galaxies, estimate their radial metallicity gradients and produce homogeneous metallicity maps.}
   {We use the XP spectra from \textit{Gaia} Data Release 3 to calculate synthetic Str\"{o}mgren magnitudes from the application of the GaiaXPy tool and adopt calibration relations from the literature to estimate the photometric metallicities.}
   {We present photometric metallicity maps for $\sim$90\,000 young stars and $\sim$270\,000 old stars within $\sim$11 deg of the Small Magellanic Cloud and $\sim$20 deg of the Large Magellanic Cloud from a homogeneous dataset. We find that the overall radial metallicity gradients decrease linearly in agreement with previous studies. Thanks to the large stellar samples, we apply piecewise-regression fitting to derive the gradients within different radial regions. The catalogues containing the estimated photometric metallicities from this work are made available.}
% conclusions (optional)
    {The overall metallicity gradients, traced by young and old stars, decrease from the centre to the outskirts of both galaxies. However, they show multiple breakpoints depicting regions following different and sometime opposite trends. These are associated to the structure of the galaxies, their history of star formation and chemical evolution but may be influenced by a low number of sources, especially at the centre (due to crowding) and in the outermost regions.}

   \maketitle

\section{Introduction}
\label{intro}

Galaxies are a multi-component (bar, bulge, disc, spiral arms, halo made of baryonic and dark matter) diverse class of objects with distinct structural, kinematical and chemical properties. They are found both in isolation as well as in groups and clusters. Morphologically, they have been classified as ellipticals, spirals and irregulars. The low mass and less luminous counterparts of these are classified as dwarfs, which are the most abundant type of galaxies in the Universe. Observationally, groups and clusters of galaxies are ubiquitous. A cluster is dense, populous and typically consists of a few tens to hundreds of galaxies bound by gravity. Whereas, a galaxy group consists of a few massive galaxies surrounded by many satellites, mostly dwarfs that have not yet dissolved or merged with their host galaxy. In the environment of galaxy clusters and groups, dynamical processes like tidal and ram-pressure stripping play a vital role in driving galaxy evolution. The $\Lambda$ cold dark matter model suggests that the dark matter halos grow hierarchically (bottom-up scenario); that is, larger halos are formed by the merging of smaller ones \citep{Fall1980,vandenbosch2002, Agertz2011}.  A detailed exploration of how this physical process affects the host and satellite galaxy is required to understand galaxy evolution in general. A system involved in both dwarf-dwarf interactions and interacting with its host (more massive galaxy) is then an excellent place to explore the implications of both interactions.\\ 
\indent One such pair of interacting galaxies is the Large Magellanic Cloud (LMC) and the Small Magellanic Cloud (SMC), which are two prominent satellites of the Milky Way (MW). They are both gas-rich dwarf irregulars and are collectively known as the Magellanic Clouds. The LMC is characterised by an inclined disc, a major spiral arm and an off-centred bar (\citealp{bekki2009,subramaniam2009}) along with evidence of warps, (\citealp{olsen2002,choi2018,saroon2022}). It is located at a distance of 50$\pm$2 kpc (\citealp{degrijs2014lmc}) in very close proximity of the SMC (62$\pm$1 kpc; \citealp{degrijs2015smc}). The SMC is characterised by an irregular shape, a wing towards the LMC and a large line-of-sight depth (\citealp{oliveira2013,Scowcroft2016,Jacyszyn-Dobrzeniecka2016,jacyszyn2017,Ripepi2017,muraveva2018}). Low-density stellar structures have been identified at the galaxy's front (Leading Arm; \citealp{putman1998}) and trailing (Magellanic Stream; \citealp{putman2003}) ends as well as between the LMC and SMC (Magellanic Bridge). These features are also prominent in \ion{H}{i} maps \citep{putman2003}. In addition, several stellar sub-structures have been found using various tracers which exhibit themselves as signatures of dynamical interactions (\citealp{mackey2016,mackey2018,elyousaffi2019,El_Youssoufi_2021,omkumar,dizna2021,almeida2024,massana2024}). Some of these sub-structures have also been associated with the influence of the MW. Hence, the Clouds serve as an excellent laboratory to study dwarf-galaxy interactions utilising resolved stellar populations. 

We expect the stellar structures that formed during the origin of the Bridge ($\sim$300 Myr ago) and Stream ($\sim$1.5 Gyr ago) to show old stars stripped from the galaxies through dynamical interactions. Previous studies \citep{nidever2011,bagheri2013,noel2013,skowron2014,noel2015,belokurov2017,jacyszyn2017,jacyszyn2019, Massana2020} found intermediate-age/old stars ($>$ 2 Gyr old) around the Bridge, but the interpretation of their location differs. \cite{noel2013,noel2015} and \cite{carrera2017} support a tidal origin for the presence of intermediate-age stars, \cite{jacyszyn2017} and \cite{wagner2017} suggest them as part of the overlapping stellar halos of the Clouds. Since the Stream is vastly spread in the sky, it is not trivial to identify any star associated with a tidally stripped population. Some Stream debris was discovered by \cite{chandra2023} and more recently, a sample of about 40 very-metal-poor stars were tentatively associated to the Stream \citep{viswanathan2024}.

Metallicities and abundance estimates are key parameters to knowing the chemical composition and also hint at the formation history and evolution of galaxies. Recent studies have significantly advanced our understanding of the metallicity distribution in Clouds using both photometry and spectroscopy. Photometric metallicity maps of the LMC ($\sim$ 4 deg from the centre; \citealp{choudhury2016lmc}) and the SMC ($\sim$ 2.5 deg from the centre; \citealp{choudhury2018ogle}) are produced using the slope of the red giant branch (RGB) as an indicator of the average metallicity of a sub-region and are calibrated using spectroscopic data. Building upon this previous work, another study extended the analysis to a larger area of the SMC ($\sim$ 4 deg from the centre; \citealp{choudhury2020smcvmc}) and the LMC ($\sim$ 5 deg from the centre; \citealp{choudhury2021lmc}). \cite{grady2021} chemically mapped the entire Clouds using data from \textit{Gaia} Data Release 2 (DR2). They utilised machine learning methods and obtained photometric metallicity estimates for the selected RGB stars using the spectroscopic metallicities from Apache Point Observatory Galactic Evolution Experiment (APOGEE) as a training samples. More recently, \cite{frankel2024} used \textit{Gaia} DR3 data to construct mono-age and mono-abundance maps of the LMC, while \cite{li2024} used the tip of the RGB stars, which has less sensitivity to interstellar reddening, to create metallicity maps of the Clouds.
 
Spectroscopic metallicities are available for only a few thousand giant stars and even less for other young populations. These measurements were also obtained using various facilities, i.e., from different instruments with varying spectral resolutions that may collectively introduce systematic uncertainties in the study of metallicity distributions. For example, \cite{dobbiefeh} estimated the metallicity gradient from the observation of RGB stars to be --0.075$\pm$0.011 dex deg$^{-1}$ out to 5 deg from the centre of the SMC. \cite{choudhury2020smcvmc} obtained a gradient of --0.031$\pm$0.005 dex deg$^{-1}$ in the inner 2.5 deg region, flattening to 4 deg. \cite{debortoli2022} investigated the metallicities of SMC stellar clusters and surrounding field stars finding that there is a bimodal distribution, a metal-poor and metal-rich group of clusters contrary to an unimodal metallicity for the SMC field stars. In addition, various studies have used different calibration relations to estimate  metallicities of similar and other stellar tracers in the Clouds \citep{cioni2009, feast2010, narloch2021smc, narloch2022}. In general, there is a lack of homogeneous and spatially extended metallicity samples that represent both the young and old stellar populations of the Clouds. These samples are essential for studying the sub-structural features in the outskirts, which will help us to understand their origins and mutual association.

\textit{Gaia} DR3 (\citealt{gaiacollaboration2016}, \citealt{montegriffo2022}, \citealt{recioblanco2022_rvs}) provides low-resolution (R=20--80) spectrophotometry for around 220 million sources, in the ranges 330--680 nm (BP) and 640--1050 nm (RP), which together are referred to as XP. A recent study by \cite{reneandrae2023} used these spectra to obtain data-driven stellar metallicities of $\sim$175 million sources including sources in the Clouds. They estimated the metallicities of the stars using the XGBoost algorithm utilising the infrared photometric data from the  ALLWISE\footnote{\href{https://wise2.ipac.caltech.edu/docs/release/allwise/}{https://wise2.ipac.caltech.edu/docs/release/allwise/}} programme and \textit{Gaia} parallaxes by training their algorithm also on the APOGEE sample. In their work, the addition of the parallaxes as one of the input parameters improved the metallicity estimates by $\sim$10\%. They presented a vetted sample with just the RGB stars after applying quality cuts mainly using the parallax values to remove the most distant sources especially those in the Clouds. The potential of this full-spectrum fitting method will further improve with subsequent data releases from \textit{Gaia}, through fixing the systematics in the spectra and aspects of the modelling. Another way of estimating the stellar metallicities from the XP spectra is by using synthetic photometry, where the transmission curve of the chosen photometric bands are completely covered by the \textit{Gaia} DR3--XP realm, to get the magnitudes and colour indices from GaiaXPy and then using relevant calibration relations to estimate the photometric metallicities. This has also been demonstrated in \cite{montegriffo2022}. Although this is an indirect way of inferring the metallicities of stars, it does have advantages over traditional spectroscopic metallicities. It is less time-consuming and hence the metallicity estimates can be made for a large sample using the same method. \cite{bellazzini2023} used this method and the available calibration relations from the literature to estimate photometric metallicities for 694\,233 Galactic giant stars from \textit{Gaia} DR3 synthetic Str\"omgren photometry. The advantage of this method is that it can also be expanded and applied to young stars, such as supergiant stars, by using appropriate calibration relations from the literature to estimate their metallicities. This is especially useful in the case of the Clouds where we have stellar populations of different ages and where a comparison of their metallicities can inform us about the chemical enrichment process within the galaxies.

In this work, we utilise the homogeneous data sample from the \textit{Gaia} DR3--XP spectra, encompassing the entire LMC and SMC, to estimate their photometric metallicities ([Fe/H]) of both young and old stars therefore, by expanding the method utilised by \cite{bellazzini2023}. We compare these metallicities with the APOGEE estimates to validate our method and we also estimate the metallicity gradients. In Section \ref{data}, we provide details on the selection of our data sample. In Section \ref{synphot}, we discuss the synthetic photometry method and in Section \ref{fehestimation}, we explain the estimation of the photometric metallicities. In Section \ref{results}, we present our results and we discuss their interpretation in Section \ref{summary}, which concludes our study.

\section{Data}
\label{data}

We describe below our initial selection of the most probable SMC and LMC sources that will be used in our study. As we aim to estimate the photometric metallicities of both the RGB and supergiant stars, we also show their further selection from the entire sample.
 
\begin{figure*}
    \centering
        \includegraphics[scale=0.6]{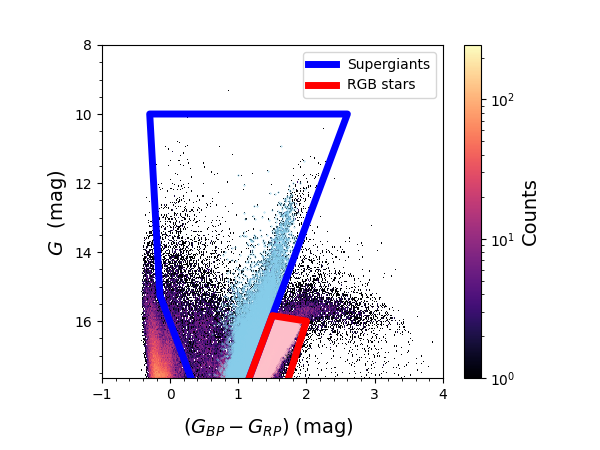} 
        \includegraphics[scale=0.6]{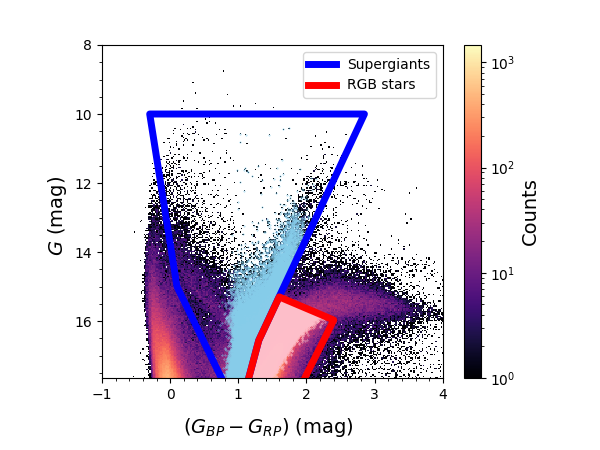} \\
        \includegraphics[scale=0.55]{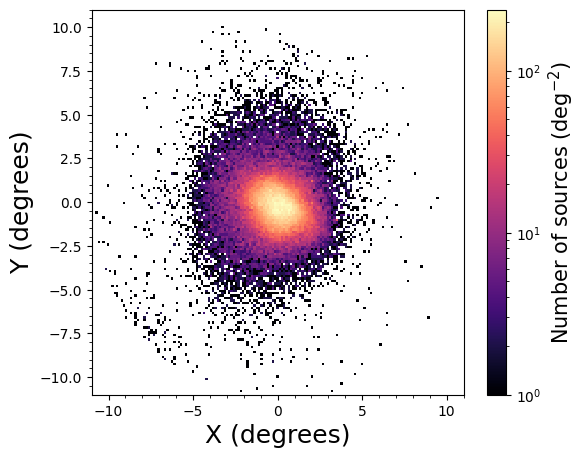}
        \hspace{1cm}
        \includegraphics[scale=0.55]{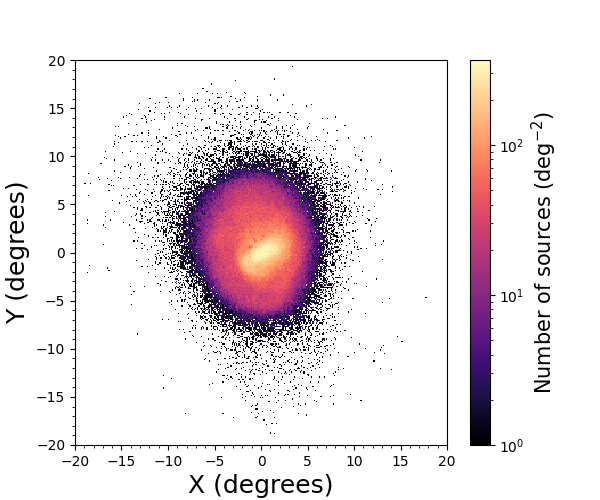}\\
        \includegraphics[scale=0.55]{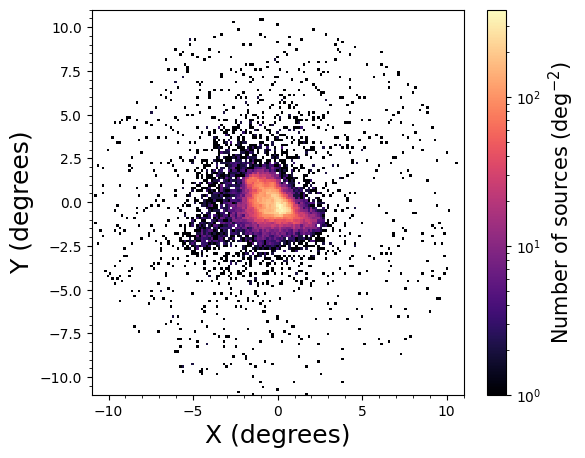} 
        \hspace{1cm}
        \includegraphics[scale=0.55]{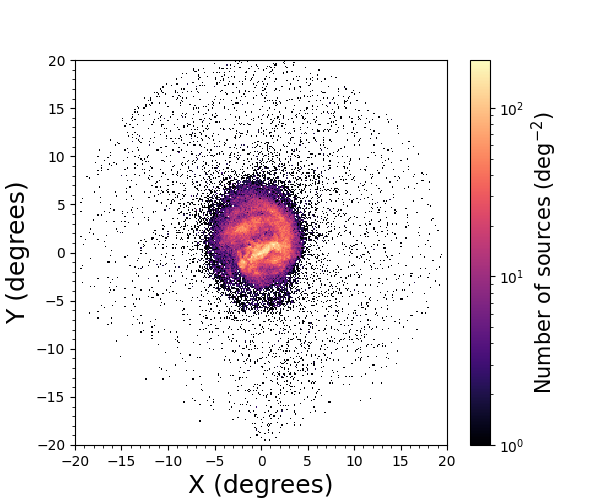} 
      \caption{\textit{Gaia} DR3 CMD of the SMC (top-left) and the LMC (top-right) sources. In both plots, the regions used to select the supergiants (blue) and the RGB stars (red) are marked. The respective final selections (see text for details) of the RGB stars (pink) and supergiants (light blue) of the Clouds are also over-plotted. The middle-left and bottom-left show the number density distribution of the selected RGB and supergiant sources within $\sim$11 deg of the SMC from its centre (RA = 12.80 deg and Dec = --73.15 deg; \citealp{cioni2000a}). The middle-right and bottom-right panels show the distribution of the RGB and supergiant sources within $\sim$20 deg of the LMC from its centre (RA = 81.24 deg and Dec = --69.73 deg; \citealp{vandermarel2001}). The distributions are shown in zenithal equidistant projection coordinates X and Y, as defined in \cite{vandermarel2001}. For the LMC we show a de-projected spatial distribution (refer to Section \ref{data_lmc}). The colour bar from black to yellow represents the increasing stellar densities in all the plots.}    
    \label{fig:smcxy}
\end{figure*}

\subsection{Gaia DR3--XP spectra for SMC sources}
\label{data_smc}
Our initial SMC sample is selected to cover a circular area with a radius of $\sim$11 deg from the SMC centre (RA = 12.80 deg and Dec = --73.15 deg, \citealp{cioni2000a}) and to include sources with magnitudes G $<$ 20.5 from the \textit{Gaia} archive\footnote{\href{https://gea.esac.esa.int/archive/}{https://gea.esac.esa.int/archive/}}. This resulted in 4\,709\,622 sources which are further reduced to 761\,736 after removing sources without \textit{Gaia} XP spectra. To further eliminate the contamination from the foreground (MW) sources and only use the most probable SMC sources, we apply the additional criteria on parallax and proper motions by following the selection procedure described in \cite{luri2021} and retain 158\,255 sources which satisfy the selection criteria. We also applied a $5\,\sigma$ cut to the flux-signal-to noises in all three \textit{Gaia} bands, resulting in sources with magnitude errors $\le$ 0.22 mag, where magnitude error = 1.086 / flux\_over\_error (\textit{Gaia} DR2 primer\footnote{\href{https://www.cosmos.esa.int/documents/29201/1773953/Gaia+DR2+primer.pdf/a4459741-6732-7a98-1406-a1bea243df79?t=1575906638431}{www.cosmos.esa.int/documents}}, Section 6). We further selected sources which have astrometric excess noise values $\le$ 1.3 mas \cite[Section~2]{omkumar}. We also estimated the photometric excess factor correction (c$_*$) by following the method suggested in the Appendix B of \cite{Gaiaedr3}, and selected stars that pass a $2\,\sigma$ criteria \cite[Equation~18]{Riello2021_excessflux}. After all these quality filters, we obtained 156\,428 sources. The RA and Dec of these sources are then converted into the zenithal equidistant projection coordinates X and Y using the transformation equations in \cite{vandermarel2001}. 

Our SMC sample consists of stellar populations of different ages as shown in the colour-magnitude diagram (CMD) on the top-left panel of Figure \ref{fig:smcxy}. As stellar populations with different ages might have different metallicities from one another, we further distinguish the RGB stars from the supergiants in our initial sample. RGB stars in the SMC are old/intermediate-age stars typically $>$ 3 Gyr old (e.g.\,\citealt{Rubele2018}) and one of the most abundant and also homogeneously distributed stellar population across the galaxy. Although we do not have spectra for the entire sample, as the faint limit of the \textit{Gaia} DR3--XP spectra is $\sim$17.65 mag, we can study the majority of bright RGB sources. To select these RGB stars, we utilise the selection criteria provided by \cite{luri2021} which define polygonal regions in the CMD occupied by different types of sources. The bright RGB sources are enclosed within the area defined by these CMD vertexes [G$_{BP}$--G$_{RP}$, G]: [0.65, 20.50], [0.80, 20.00], [0.80, 19.50], [1.60, 19.80], [1.60, 19.60], [1.00, 18.50], [1.50, 15.843], [2.00, 16.00], [1.60, 18.50], [1.60, 20.50] and [0.65, 20.50]. Further, we select the supergiant stars. The typical age range of supergiants in the Clouds is 30--250 Myr (e.g.: \citealp{luri2021}). By using the following polygon selection, we separated our supergiant stars which fall in the Blue Loop region from the entire SMC sample, [G$_{BP}$--G$_{RP}$, G]: [0.40, 18.15], [--0.15, 15.25], [--0.3, 10.00], [2.60, 10.00], [1.00, 18.50], [0.80, 18.50] and [0.40, 18.15]. Finally, there are 78\,833 RGB stars and 39\,324 supergiants in our sample. Their spatial distribution is shown in the middle- and bottom-left panels of Figure \ref{fig:smcxy}, respectively.

\subsection{Gaia DR3--XP spectra for LMC sources}
\label{data_lmc}
A similar procedure to the one adopted for the SMC in the previous section is followed to select the LMC sources but using a 20 deg radius from its centre (RA = 81.24 deg and Dec = --69.73 deg; \citealp{vandermarel2001}). Our base LMC sample consists of 27\,231\,400 objects, which includes stellar populations of different ages as shown on the top-right panel of Figure \ref{fig:smcxy}. Since there are many objects in the LMC and our interest lies in the RGB and supergiant stars, we initially apply the selection criteria provided by \cite{luri2021} for extracting these two populations from the CMD. To select LMC RGB sources we use as vertexes of the polygon [G$_{BP}$--G$_{RP}$, G]: [0.80, 20.5], [0.90, 19.5], [1.60, 19.8], [1.60, 19.0], [1.05, 18.41], [1.30, 16.56], [1.60, 15.3], [2.40, 15.97], [1.95, 17.75], [1.85, 19.0], [2.00, 20.5] and [0.80, 20.5] and to select LMC supergiants we use instead [G$_{BP}$--G$_{RP}$, G]: [0.90, 18.25], [0.1, 15.00], [--0.30, 10.0], [2.85, 10.0], [1.30, 16.56], [1.05, 18.41] and [0.90, 18.25]. The panels on the middle- and bottom-right of Figure \ref{fig:smcxy} show the distribution of the selected 520\,338 RGB and 99\,323 supergiant stars of the LMC, respectively.

The LMC is an inclined disc galaxy and it is important to de-project the spatial distribution from the sky plane to the galaxy plane. Hence the inclination (\textit{i}) of the LMC disc with respect to the sky plane and the position angle of the line of nodes ($\theta$) are taken into account to estimate the de-projected X$'$ and Y$'$ coordinates. We apply the recent estimates of \textit{i} = 23.26 deg and $\theta$ = 160.43 deg from \textit{Gaia} Early DR3 \citep{saroon2022} and used the distance estimate of $\sim$50 kpc from \cite{degrijs2014lmc}. The de-projected spatial distribution is used hereafter for the LMC during the entire analysis.

\section{Synthetic Str\"{o}mgren photometry using \textit{Gaia} DR3--XP spectra}
\label{synphot} 

Bengt Str\"{o}mgren introduced a narrow-band photometric system with four (\textit{u}: 350 nm, \textit{v}: 410 nm, \textit{b}: 467 nm and \textit{y}: 547 nm) bandpass filters (refer to \citealt{stromgren1963,stromgren1964} for more details) covering a similar range as the UBV photometric system. These narrow Str\"{o}mgren bands are intentionally designed to match the specific signatures in the spectra of stars that directly relate to the physical properties Teff, log g and [Fe/H] and therefore allow us to classify stars of different spectral classes 
(\citealt{stromgren1964}). Subsequent studies e.g.: 
\cite{grebelrichtler1992}, built on the photometric system devised by Str\"{o}mgren and derived calibration relations to estimate photometric metallicities of RGB stars 
(\citealt{hilker2000,calamida2007,hughes2014,narloch2021smc,narloch2022,bellazzini2023} and references therein) and supergiants 
(\citealt{grebelrichtler1992,arnadottir2010,piatti2019}). These studies provided metallicity estimates also for stars in the Clouds and of star clusters. 

\cite{montegriffo2022} and \cite{deangeli2022} explored the details of \textit{Gaia} DR3--XP spectra and also the generation of synthetic photometry in bands that are covered by the \textit{Gaia} BP/RP wavelength ranges. However, the residual systematics in the externally calibrated XP spectra have larger discrepancies when compared with the existing photometry notably at $\lambda$ $<$ 400 nm.  \cite{Giacomo2023} also produced synthetic photometry in BVI bands from the \textit{Gaia} DR3--XP spectra and validated the accuracy of the synthetic photometric conversion using GaiaXPy by comparing it with literature values. The GaiaXPy is a Python tool from the Data Processing and Analysis Consortium (DPAC) that allows the generation of synthetic photometry in a set of desired systems from the input internally calibrated continuously-represented mean spectra (see \citealt{montegriffo2022} for more details). \cite{montegriffo2022} used a second-level calibration called standardisation to reproduce existing photometry in several widely used systems to millimag accuracy. Section 4 of 
\cite{montegriffo2022} discusses the potential of XP spectra by estimating the synthetic Str\"{o}mgren magnitudes, as a test case, and deriving photometric metallicities. In our study, we follow a similar method. First, we query the \textit{Gaia} archive to obtain the \textit{Gaia} DR3--XP spectra for a list of probable SMC and LMC sources. Then, we use the GaiaXPy tool to derive the standardised synthetic Str\"{o}mgren magnitudes (\textit{v}, \textit{b} and \textit{y}). Note that the standardisation is not done for the u band and the calibration relations we choose do not utilise it hence, we do not use it in our analysis. GaiaXPy also provides us with fluxes and flux errors on each of the Str\"{o}mgren bands.

\section{Estimation of photometric [Fe/H]}
\label{fehestimation}
\begin{figure*}
	\centering
	\includegraphics[scale=0.3]{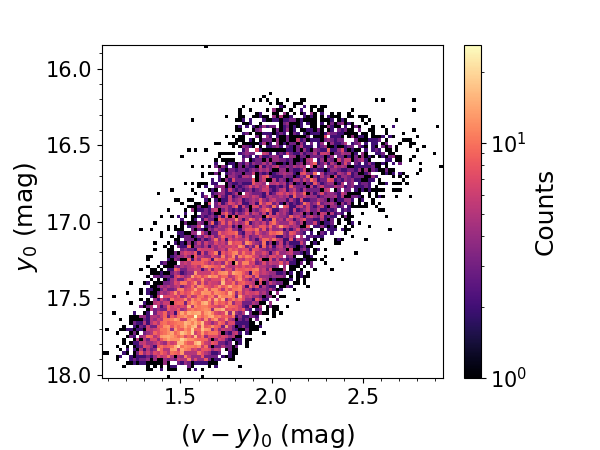}\hfill 
	\includegraphics[scale=0.3]{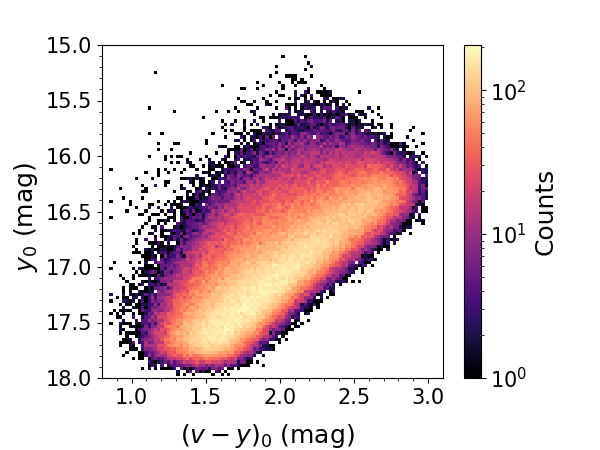}\hfill
	\includegraphics[scale=0.3]{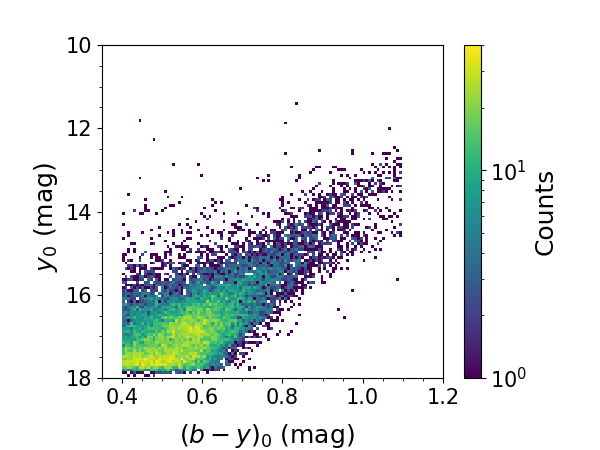}\hfill
	\includegraphics[scale=0.3]{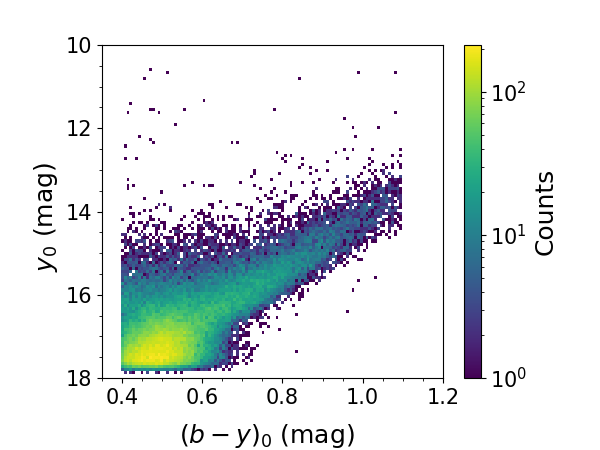}  
	\caption{The CMDs depict the final selection of RGB stars in the SMC (first plot) and those in LMC (second plot), both using Str\"{o}mgren bands. The third and fourth plots present the CMDs of the final selection for supergiants in the SMC and in LMC, also using Str\"{o}mgren bands.}
	\label{fig:cmd}
\end{figure*}

\subsection{Calibration relation for RGB stars}
\label{rgbfeh}

In a recent study \cite{bellazzini2023} estimated the photometric metallicities of Galactic giant stars by utilising synthetic Str\"{o}mgren magnitudes from \textit{Gaia} DR3--XP spectra. Their metallicities derived from the empirical calibration relations provided by \cite{calamida2007}, reported in equation (\ref{equation:calamidafeh}), have a typical accuracy of $\le$ 0.1 dex in the range --2.2 < [Fe/H] < --0.4 dex. They also found a systematic trend with [Fe/H] at higher metallicity, beyond the applicability range of the calibration relation. Metallicities derived using the \cite{hilker2000} calibration are less precise, and show particularly lower accuracy in the metal-poor regime when compared with the spectroscopic metallicities from APOGEE. 

	{\begin{equation}
			\text{[Fe/H]} = \frac{(m_{1,0} - \gamma * (v - y)_0 - \alpha)}{\delta * (v - y)_0 + \beta}
			\label{equation:calamidafeh},
		\end{equation}
		where  $\alpha$ = --0.312, $\beta$ = --0.096 $\pm$ 0.002, $\gamma$ = 0.513 $\pm$ 0.001, $\delta$ = 0.154 $\pm$ 0.006 and $m_{1} = (v-b) - (b-y)$.\\
}

In this study, we use the \cite{calamida2007} calibration to calculate the iron abundance [Fe/H] of the SMC and LMC RGB stars. To correct for extinction, we use the reddening maps of the Clouds provided by \cite{skowron2021} where the extinction is estimated using the red clump giant stars. Their maps cover areas corresponding to a distance from the centre of the galaxies of 2.5 deg for the SMC and 6.5 deg for the LMC. In the outer regions, that are not covered in \cite{skowron2021} maps, the extinction is corrected by using the all-sky reddening map of the MW provided by \cite{schlegel1998}. The maps provide median $E(B-V)$ values and for each of the sources in our study we use the nearest region for getting their extinction values. The reddening values $E(V-I)$ provided by \cite{skowron2021} can be converted into $E(B-V)$ by using the recalibration by \cite{schlafly2011}. As the recalibration suggests, we translated the $E(V-I)$ into $E(B-V)$ by $E(B-V) = (E(V-I)/1.237)\times0.86$. 
We obtain the dereddened colour indices, $m_{1,0} = m_1 + 0.24 \times E(B-V)$, $(v-y)_0 = (v-y) -1.24 \times E(B-V)$, and $(b-y)_0 = (b-y) -0.74 \times E(B-V)$ using the equation from \cite{crawford1976}. 

The applicability range of equation (\ref{equation:calamidafeh}) is $0.85 \le (v-y)_0 \le 3$ which reduces our SMC and LMC RGB samples to 78\,473 and 512\,367 sources, respectively. Then, using equation (\ref{equation:calamidafeh}), we estimate the photometric metallicities of the SMC and LMC RGB stars finding a significant spread. On further investigation, it is clear that this is due to those sources with magnitude uncertainties in each band $>$ 0.1 mag; eliminating them results in 57\,512 SMC and 214\,603 LMC RGB stars. The CMDs in Figure \ref{fig:cmd} clearly illustrate the range of $(v-y)_0$ colours of our selected RGB stars in the SMC (first plot) and LMC (second plot) using Str\"{o}mgren bands, which are also over-plotted in the \textit{Gaia} CMDs (Figure \ref{fig:smcxy}). We follow the error propagation to estimate the uncertainties on the metallicities and the estimated median error of both samples is around 0.6 dex. In Figure \ref{fig:fehhist}, the distribution of the estimated metallicities (top-left) and the estimated errors (bottom-left) for both galaxies are provided. The estimation of [Fe/H] involved multiple quantities all of which got propagated which resulted in large uncertainties. The distribution shows that the majority of the sources have uncertainties $<$1 dex with on average larger values for supergiants than for RGB stars. We did not
limit the uncertainties of the metallicities in our study which is then based on the full sample.

\subsection{Calibration relation for supergiant stars}
\label{supergiantfeh}

To estimate the metallicities of the supergiant stars we use the empirical relations provided by \cite{grebelrichtler1992} and reported in equation (\ref{equation:feh}). They successfully performed a feasibility study to apply this equation on the supergiants in the young stellar cluster NGC 330 and also on some field stars of the SMC around the cluster. Later, a study by \cite{piatti2019} applied this relation to estimate the metallicities of the supergiant stars in selected young clusters of the Clouds.

\begin{equation}
    \text{[Fe/H]} = \frac{(m_{1,0} + a_1 * (b - y)_0 + a_2)}{a_3 * (b - y)_0 + a_4},
    \label{equation:feh}
\end{equation}

where $a_1 = -1.240$$\pm$0.006, $a_2 = 0.294$$\pm$0.030, $a_3 = 0.472$$\pm$0.040 and $a_4 = -0.118$$\pm$0.020.

\begin{figure*}
    \centering
     \includegraphics[width=1\columnwidth]{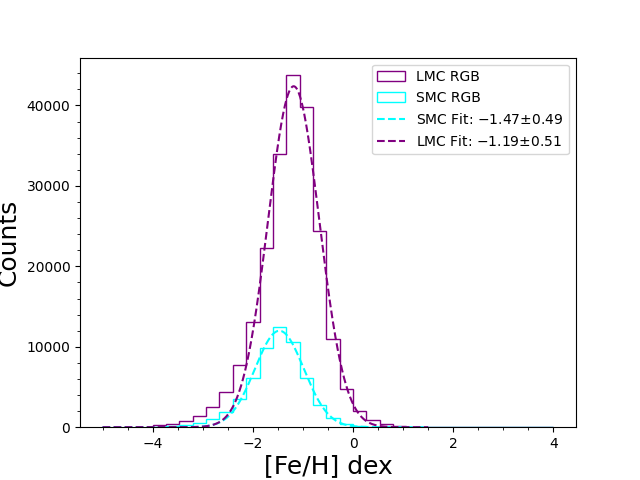} 
     \includegraphics[width=1\columnwidth]{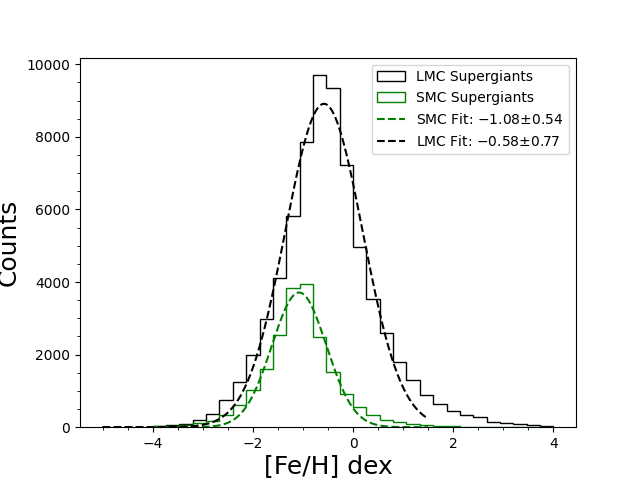} \\
     \includegraphics[width=1\columnwidth]{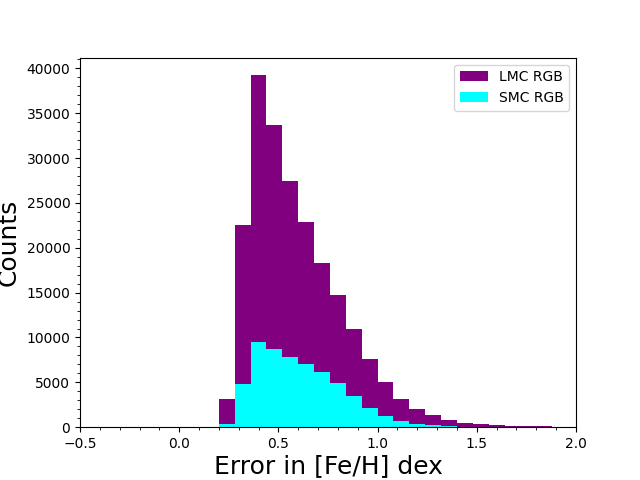} 
     \includegraphics[width=1\columnwidth]{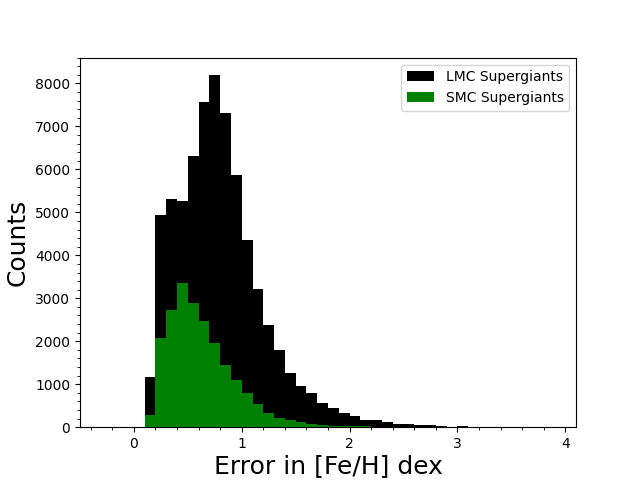} \\
     \caption{Metallicity distributions of RGB (top-left) and supergiant (top-right) stars of the LMC and SMC samples. Best-fit Gaussians for the SMC (cyan) and the LMC (purple) are shown on the histograms. The distributions of the estimated uncertainties are also shown for RGB stars (bottom-left) and for supergiants (bottom-right). Best-fit Gaussians for the SMC (green) and the LMC (black) are marked.}
    \label{fig:fehhist}
\end{figure*}

We followed the same method as explained in Section \ref{rgbfeh} to estimate the reddening values. Then, we estimate the extinction values in the visual band $A_V = 3.1\times E(B-V)$. We use the translation equation from \cite{cardelli1989} and we estimate the extinction in each Str\"{o}mgren band by using $A_v = 1.397\times A_V$, $A_b = 1.240\times A_V$ and $A_y = 1.005\times A_V$. The applicability range of equation (\ref{equation:feh}) is $0.4 \le (b-y)_0 \le 1.1$ and this reduces the SMC sample to 21\,207 sources and the LMC sample to 71\,628 sources. Using equation (\ref{equation:feh}) we estimate the photometric metallicities of the supergiant sources in both galaxies. Similarly to the RGB stars, we remove sources with magnitude uncertainties in each band $>$ 0.1 mag, which results in 20\,713 SMC and 69\,083 LMC supergiants. Figure \ref{fig:cmd} shows the range of $(b-y)_0$ colours for our selected supergiant stars in the SMC (third plot) and LMC (fourth plot) using the Str\"{o}mgren bands. The distribution of the estimated metallicities of the SMC and LMC supergiants are shown on the top-right panel of Figure \ref{fig:fehhist}. We also estimate the uncertainties for our [Fe/H] values using error propagation and obtained a median error of around 0.8 dex for the SMC and around 0.6 dex for the LMC samples. The propagated error distribution for both galaxies are shown in the bottom-right panel of Figure \ref{fig:fehhist}. As for RGB stars, no further reduction has been applied based on the errors. 

We chose to utilise the \cite{grebelrichtler1992} calibration for supergiants and the \cite{calamida2007} for our RGB samples for the following reasons. \cite{calamida2007} offers a new empirical metallicity calibration using $(u-y)$ and $(v-y)$ colours, which are more sensitive to temperature across a broader metallicity range of --2.2 <= [Fe/H] <= --0.7 dex, providing an advantage over \cite{grebelrichtler1992}. Additionally, \cite{bellazzini2023} successfully applied the \cite{calamida2007} calibration for estimating metallicity in Galactic giant stars, yielding accurate results. However, \cite{montegriffo2023} noted that this calibration, based on older globular clusters, may introduce systematic offsets when applied to younger systems, like open clusters, due to intrinsic age differences. In contrast, despite its limited sources, the \cite{grebelrichtler1992} calibration is applicable to red supergiants, making it more suitable for our purpose of estimating metallicities for these stars.

\section{Results}
\label{results}

\subsection{Metallicity distributions}
\label{feh_distribution}

The histograms in the top-left and top-right panels of Figure \ref{fig:fehhist} depict the metallicity distributions of the RGB stars and supergiants of the Clouds, respectively. We excluded sources with [Fe/H] and uncertainties outside the broad range illustrated in Figure \ref{fig:fehhist}, resulting in 271\,843 RGB stars and 89\,599 supergiants in the Clouds after removing duplicate sources in the outskirts. By comparison, the overall metallicity of the SMC RGB stars is lower than that of the LMC as expected, which is also the case for the supergiants. For both populations, we obtain unimodal metallicity distributions in agreement with the previous studies \citep{carrera2008,dobbiefeh,choudhury2015,grady2021,debortoli2022}. These literature metallicity distributions also show significant asymmetries, with sharp declines towards the metal-rich ends and extended tails towards the metal-poor ends (see e.g. \citealp{grady2021}). We do not see these trends in our metallicity distributions, probably because of the large errors on the individual metallicity estimates (Figure \ref{fig:fehhist}). We recognise that while the shape of our metallicity distributions may have limitations due to larger uncertainties, our focus is on the mean (median) metallicities because our main goal is to produce maps of the mean metallicity and mean metallicity gradients. To further compare the metallicity distributions, we estimate the peak and dispersion values by fitting a Gaussian to each of the histograms. The peak metallicity and the dispersion of the LMC RGB stars is --1.19$\pm$0.51 dex whereas for the supergiants it is --0.58$\pm$0.77 dex. In the SMC the RGB stars show a peak metallicity and a dispersion of --1.47$\pm$0.49 dex while that of the supergiants is of --1.08$\pm$0.54 dex. There could also be larger systematic errors in these estimates, for instance caused by errors in the calibration or between the RGB stars and the supergiants.
\cite{grady2021} estimated the median metallicity of RGB stars within 12 deg of the LMC to be --0.78 dex. \cite{haschke2012rrlyrae} obtained a mean [Fe/H] of --1.50$\pm$0.24 dex based on RR Lyrae stars in the LMC. Previous studies like \cite{dobbiefeh} and  \cite{choudhury2015} found the median [Fe/H] to be about --1 dex from the analysis of RGB stars in the inner 4--5 deg of the SMC. \cite{carrera2008} also quote a similar value of --1 dex for RGB stars in the inner region, but they also note a decrease of the median value towards the outer region of the galaxy. In the SMC, \cite{haschke2012rrlyrae} obtained --1.70$\pm$0.27 dex based on RR Lyrae stars. Our estimates for both the LMC and SMC RGB stars are lower compared to the median values estimated previously. This could be due to the fact that our metallicity distributions include stars in large areas encompassing the SMC and LMC outer regions which are predominantly populated by metal-poor and old stars. However, a recalibration of our metallicities using the APOGEE data shows a systematic difference of about 0.4 dex (see Section \ref{comparison}) which would bring our estimates in line with those from previous studies. If we were to limit our sample, by applying cuts based on the individual metallicity errors (e.g. 0.3 dex), it will reduce to only a few thousand sources. Irrespective of our large uncertainties, we see that our median metallicity estimates align with those from previous studies. Hence, we did not apply further cuts in order not to reduce the statistical significance of our analysis. We examined our initial selection of photometric and astrometric cuts to assess how different criteria impact the resulting metallicity distributions. Our findings indicate that the metallicity distributions (Figure \ref{fig:fehhist}) remain unchanged when using \textit{ruwe} $\le$1.2 and applying photometric error of $<$0.1 mag instead of $<$0.22 mag as adopted in all three \textit{Gaia} bands.

\subsection{Photometric metallicity recalibration}
\label{comparison}

In order to validate the estimates of the photometric metallicities of the RGB and supergiant samples from our work, we compare them with the spectroscopic metallicities from APOGEE. APOGEE is a high-resolution (R$\sim$22\,000), near-infrared ($\textit{H}$ band; 1.51--1.7 $\mu$m) spectroscopic sky survey. In the southern sky, observations are performed with the 2.5m du Pont telescope at Las Campanas Observatory, Chile. The initial samples of sources in the LMC and SMC regions were $\sim$70\,000. Though the survey mainly focused on the RGB stars, it also included main sequence stars, asymptotic giant branch (AGB) stars and post-AGB stars. The details on the target selection are explained in \cite{zasowski2017}. The ASPCAP (APOGEE Stellar Parameters and Chemical Abundances Pipeline) calculates the stellar parameters and elemental abundances. We cross-matched the APOGEE sources with the \textit{Gaia} DR3 data, to select the most probable members of the Clouds and retain those with parallax $\le$ 0.2 mas which eliminates foreground MW sources within $\sim$5 kpc. Furthermore, we apply the proper motion selections (1 $\le$  $\mu_{\alpha}$  $\le$ $+$2.5 mas yr$^{-1}$ and $-$0.8 $\le$ $\mu_{\delta}$ $\le$ 1.5 mas yr$^{-1}$) to identify the most probable LMC sources \citep{saroon2022}. This selection criteria removes many sources and we are left with a sample of $\sim$14\,135 sources. Along with the cut in the parallax to remove the foreground contamination from the MW, we also discarded the SMC sources whose proper motions lie outside the expected range ($-$3 $\le$  $\mu_{\alpha}$, $\mu_{\delta}$ $\le$ $+$3 mas yr$^{-1}$) as predicted by the simulations \citep{Diaz2012} for the main body and stellar tidal features around the SMC. We recalibrated our photometric estimates to match the zero-points of both our samples (Appendix \ref{apogee}). This is done by crossmatching the APOGEE sample of the SMC and LMC RGB sources with our samples after the elimination of sources with large photometric uncertainties. The crossmatched sources (4308) then have both the spectroscopic metallicity estimates from APOGEE [Fe/H]$_\text{spec}$ and the photometric metallicity estimates from our study [Fe/H]$_\text{phot}$. We estimate the peak of  [Fe/H]$_\text{spec}$ and [Fe/H]$_\text{phot}$ from Gaussian fitting and computed their difference, which is about $-0.43$$\pm$0.02 dex, for both the SMC and the LMC RGB samples. This value is used to obtain the recalibrated metallicities such as [Fe/H]$_\text{recal}$ = [Fe/H]$_\text{phot}$ + 0.43 dex. \cite{bellazzini2023} also used a similar method to recalibrate the metallicity estimates of their MW sources. Following a similar procedure for supergiant stars, we found 1212 crossmatches with APOGEE. The peak difference obtained is $-0.35$$\pm$0.01 dex. Hence, the recalibrated metallicities is obtained as [Fe/H]$_\text{recal}$ = [Fe/H]$_\text{phot}$ + 0.35 dex. \cite{bellazzini2023} found a smaller shift of about 0.14 dex for MW stars. The spatial distribution of the Clouds (see Figure \ref{fig:smcxy}) indicates that there is a decreasing completeness towards the centre of the galaxies. This suggests an increased level of crowding that can lead to blending and contamination of XP spectra. This may impact our estimation of the photometric metallicities, especially at the centre of the Clouds (\citealp{Rathore_2025}). In this context, to better understand the larger shifts obtained in our study, we performed a test to determine whether the peak difference between our photometric metallicities and spectroscopic sample decreases as we move away from the centre of the Clouds (refer to Section \ref{amplitudeshift} and Figures \ref{fig:amplitude_shift_rgb} and \ref{fig:bl_amplitude_shift}). 
 
 \subsection{Radial recalibration}
 \label{rr}
To maintain consistency in our recalibration with respect to the spectroscopic sample, it is essential to account for variations in the recalibration zero point. This consideration is particularly significant in regions affected by crowding, as they will experience the most impact from radial bias when applying uniform recalibration throughout the galaxy. Our analysis of the two galaxies was conducted separately for the RGB stars. It is evident from the top two rows of Figure \ref{fig:amplitude_shift_rgb}, that the peak differences change from the centre to the outskirts. In the bottom-left panel of Figure \ref{fig:amplitude_shift_rgb}, we observe that the radial variation of the LMC RGB peak difference is significantly larger at the centre (0.7 dex) than at the outskirts. As we move outward, this difference decreases to 0.3 dex, and then slightly increases to 0.5 dex in the outer regions. However, in the outer bins, the distribution of sources becomes less clear due to the reduced number of stars present. The important factor contributing to these larger shifts could be crowding, causing an increased spectral contamination. Therefore, radial recalibration is applied to the RGB stars in the LMC. For the defined radial bins shown in Figure \ref{fig:amplitude_shift_rgb}, we utilise the corresponding peak difference rather than a constant value. In contrast, the third row of Figure \ref{fig:amplitude_shift_rgb} shows that the RGB stars in the SMC do not exhibit significant radial differences and have larger standard deviations. As a result, we opted for a consistent recalibration factor of 0.4 dex across the entire SMC.\\
 \indent We encountered limitations regarding the SMC supergiants due to an insufficient number of cross-matched sources from the APOGEE survey. This indicates that significant differences may not be detectable between our photometric and spectroscopic samples. Therefore, we did not further separate the LMC and SMC supergiants (Figure \ref{fig:bl_amplitude_shift}). The middle plot in the bottom panel of Figure \ref{fig:bl_amplitude_shift} primarily represents the SMC supergiant stars based on their radial distance. For all supergiant stars, we observed an overall constant difference of 0.4 dex, with a variation of approximately 0.1 dex only in the bin corresponding to the radial range of 3–4 kpc. Given the large standard deviations, we applied a constant recalibration factor of 0.4 dex.

\subsection{Metallicity maps of the RGB and supergiant stars}

\begin{figure*}
	\centering
	\vspace*{-0.4cm}
	\includegraphics[width=1\columnwidth]{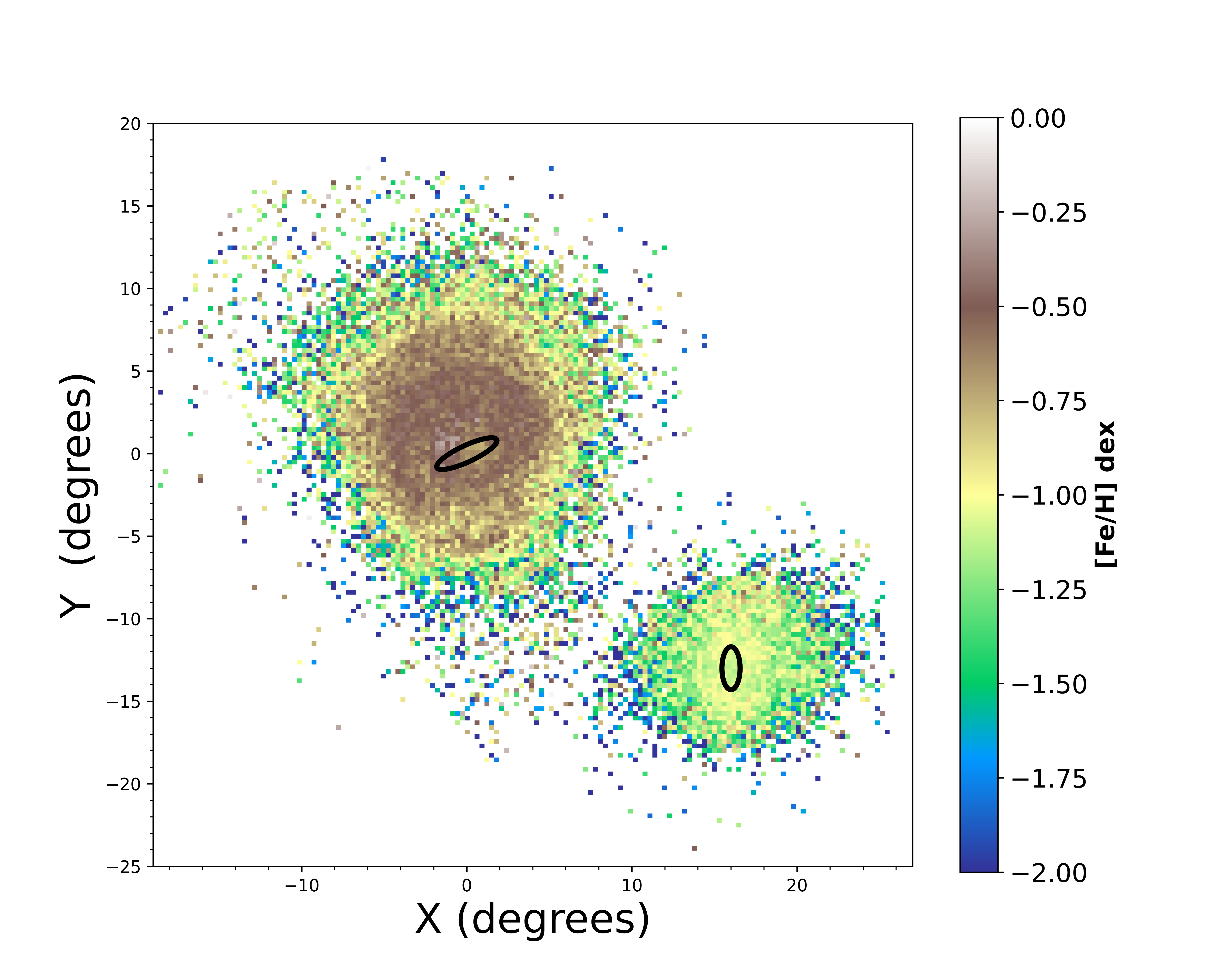}	
	\hfill
	\includegraphics[width=1\columnwidth]{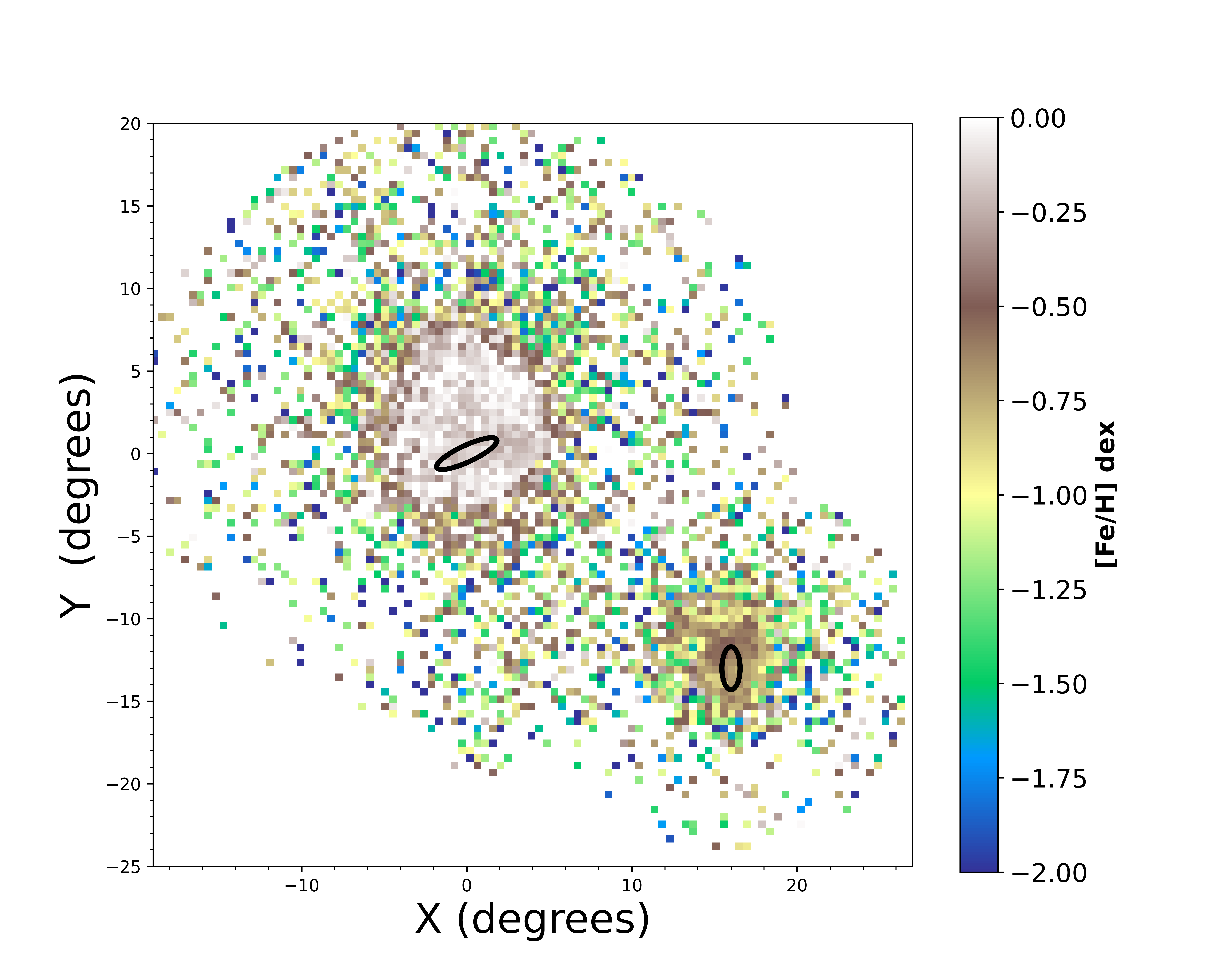} 
	\caption{Str\"{o}mgren photometric metallicity maps of RGB (left) and supergiant (right) stars of the Clouds derived from the \textit{Gaia} DR3--XP spectra. Each spatial bin corresponds to 0.25 deg$^2$ and 0.5 deg$^2$ for the RGB and supergiant stars, respectively. The maps are centred at RA = 81.24 deg and Dec = --69.73 deg; \citep{vandermarel2001}. The bar regions of both galaxies are marked with black ellipses. The colour-coding from blue to white shows the increasing median metallicity values. Note the larger relative difference between the LMC and SMC metallicity in the supergiants than in the RGB stars. The distribution of RGB stars, which contains fainter sources than that of supergiants, at the core of the galaxies may be affected by a lack of sources due to crowding.}
	\label{fig:feh}
\end{figure*}

\begin{figure*}
	\centering
	\vspace*{-0.4cm}
	\includegraphics[width=1\columnwidth]{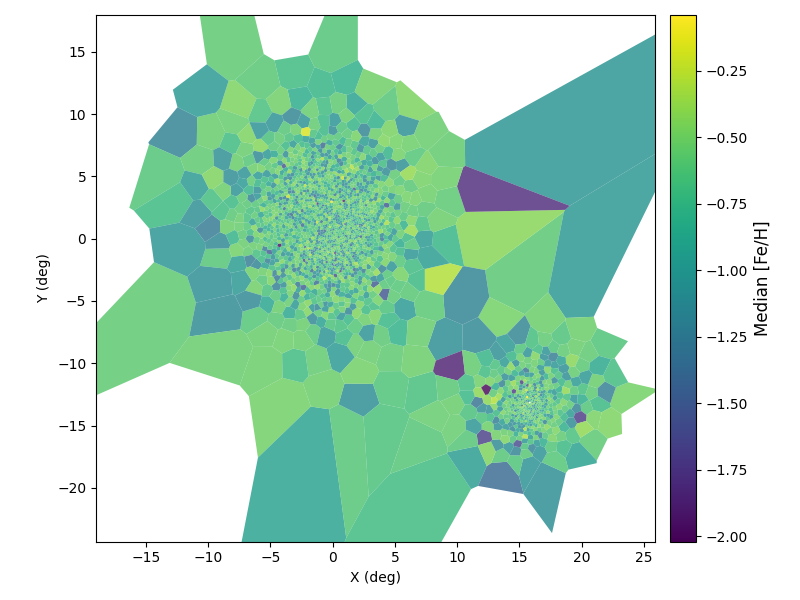}
	\hfill	
	\includegraphics[width=1\columnwidth]{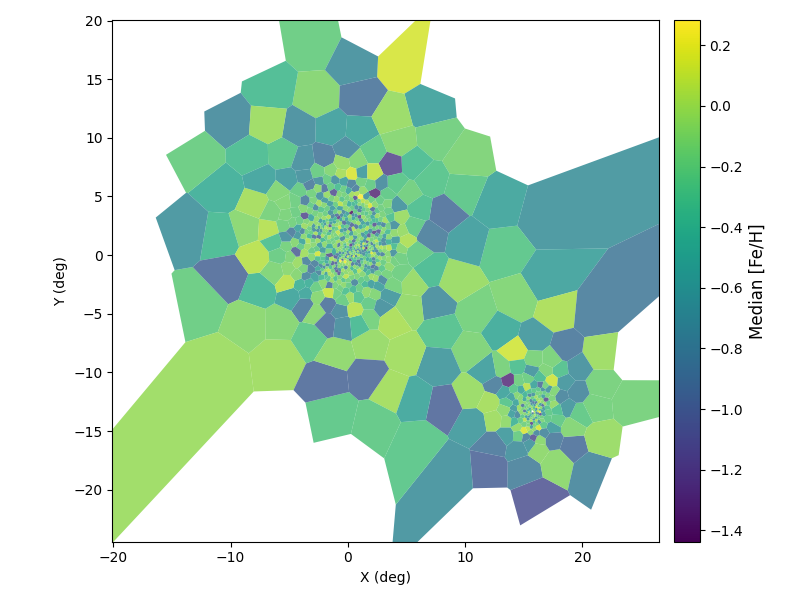}\\
	\caption{Str\"{o}mgren photometric metallicity maps of RGB (left) and supergiants (right), respectively. These maps are produced using the Voronoi binning method, where each bin has 100 sources. The colour-coding from green to yellow indicates an increase in the metallicity.}
	\label{fig:voronoi_feh}
\end{figure*}

We produce the metallicity maps for the young (supergiants) and the old (RGB stars) stellar populations separately. In order to do that we combine the RGB sources from both the SMC and the LMC and remove any duplicate sources as our initial selection of the LMC and SMC areas overlap in their outer regions. Then, we create a two-dimensional Hess distribution (a density plot) colour-coded with the median of the metallicity of each bin, using the recalibrated metallicity estimates, since the median is less prone to the effect of outliers. The left panel of Figure \ref{fig:feh} shows the metallicity map of the RGB stars where each bin corresponds to 0.25 deg$^2$. Although we tend to have fewer sources in the outskirts of the Clouds than in their inner regions, their overall distribution is a large improvement over the stellar density from previous spectroscopic metallicity studies. The central regions of the LMC reveal a metal-rich inner disc and a southern spiral arm. The bar is not obvious and we note that at its centre the metallicity is low, but this area is also influenced by a lack of sources in our sample due to crowding. The median value of each bin represents the underlying population accurately only with a sufficient number of stars, especially when individual measurement uncertainties are significant. To address this, we created metallicity maps using Voronoi binning as shown in Figure \ref{fig:voronoi_feh}. Here, we adopted equal density binning, ensuring that each bin has 100 sources, based on the median error derived from Markov Chain Monte Carlo (MCMC) simulations (refer to section \ref{voronoi} for more details). In the left panel of Figure \ref{fig:voronoi_feh} we show the metallicity distribution of RGB stars in the Clouds. The LMC's central region has a higher metallicity (yellow-green shades), while the SMC displays a scattered, lower metallicity distribution (greener to purple regions). The outer regions of both galaxies transition into lower metallicities, consistent with the expectation that older, less metal-enriched stars dominate the outskirts. This spatial variation reflects the galaxies' evolutionary histories, with stronger star formation and chemical enrichment at the centres. The gradient from higher to lower metallicity corresponds to typical dwarf galaxy evolution models, while small-scale variations signal complex star formation histories. We also created Voronoi binning plots for the median uncertainty in each bin (see Figure \ref{fig:voronoi_sigma_feh}).\\
\indent Recently \cite{frankel2024} produced a metallicity map of the entire LMC using total metallicity estimates ([M/H]) from \cite{reneandrae2023}. Their maps confirm an extended disc between --1.6 dex and --0.8 dex compared to a smaller disc at metallicities below --1.6 dex or above --0.8 dex. The bar region dominates at all metallicities, but spiral features are not as prominent as in our map. \cite{grady2021} provided the metallicity maps of the Clouds using the \textit{Gaia} DR2 data. Their maps also reveal a metal-rich LMC bar and a metal-rich SMC centre. In the outer regions, the metallicity further decreases as we see also in our maps. The right panel of Figure \ref{fig:feh} provides the metallicity map of the supergiant (young) stars of the Clouds using bins with a size of 0.5 deg$^2$ due to the reduced number of sources. Similarly to the most metal-rich maps by \cite{frankel2024}, we trace the extent of the inner LMC disc superimposed to a wide metal-poor outer region, but we do not find a clear enhancement corresponding to the bar of the galaxy. In the SMC, our map clearly traces a metal-rich bar-like feature at its centre, which is however metal poorer than at the centre of the LMC. The young populations are metal-richer than the old ones, as previously noticed (Figure \ref{fig:fehhist}) and their spatial distribution differ. A similar trend can be seen in the right panel of Figure \ref{fig:voronoi_feh}, which shows the Voronoi-based metallicity distribution of supergiants in the Clouds. Each bin has 100 stars and the colour scale indicates metallicity levels, with yellow representing high metallicity and purple indicating low metallicity. These findings align with previous research, such as \cite{grady2021}, which highlighted the LMC's inside-out formation and the SMC's bursty star formation history. Additionally, \cite{frankel2024} noted the impact of interactions between the LMC and SMC on their metallicity distributions, further explaining the observed differences. Both of these works utilised different machine learning methods, which shows dependency on their respective training samples. For example, in Figure 8 of \cite{frankel2024}, it is evident that the difference in [M/H] estimates between their work and the APOGEE DR17 sources is less than 0.2 dex for G$\le$16.5 and less than 0.4 dex at fainter magnitudes, which is closer to the 0.43 dex of constant recalibration as we obtained for the RGB sample in our study. But, the figure also includes differences from other spectroscopic surveys, which show a wider range than the APOGEE DR17 data. It is important to note that a large number of sources from APOGEE were used to train the XGBoost algorithm \citep{reneandrae2023} utilised by \cite{frankel2024}. Therefore, the observed smaller differences could be biased due to this training on APOGEE DR17 data. Both of these works demonstrate the ability of data-driven methods to estimate the photometric metallicities for a larger sample of stars but focused exclusively on the older population. This is understandable, as there is a larger amount of spectroscopic data available for the RGB stars to train such models. In contrast, our work aims to estimate the metallicities of a larger sample of stars that includes both young and old stellar populations using the same method. This approach is particularly beneficial for galaxies like the Clouds, where studying populations of different ages is essential to understanding their complex interaction history.

\begin{figure*}
    \centering
     \includegraphics[width=1\columnwidth]{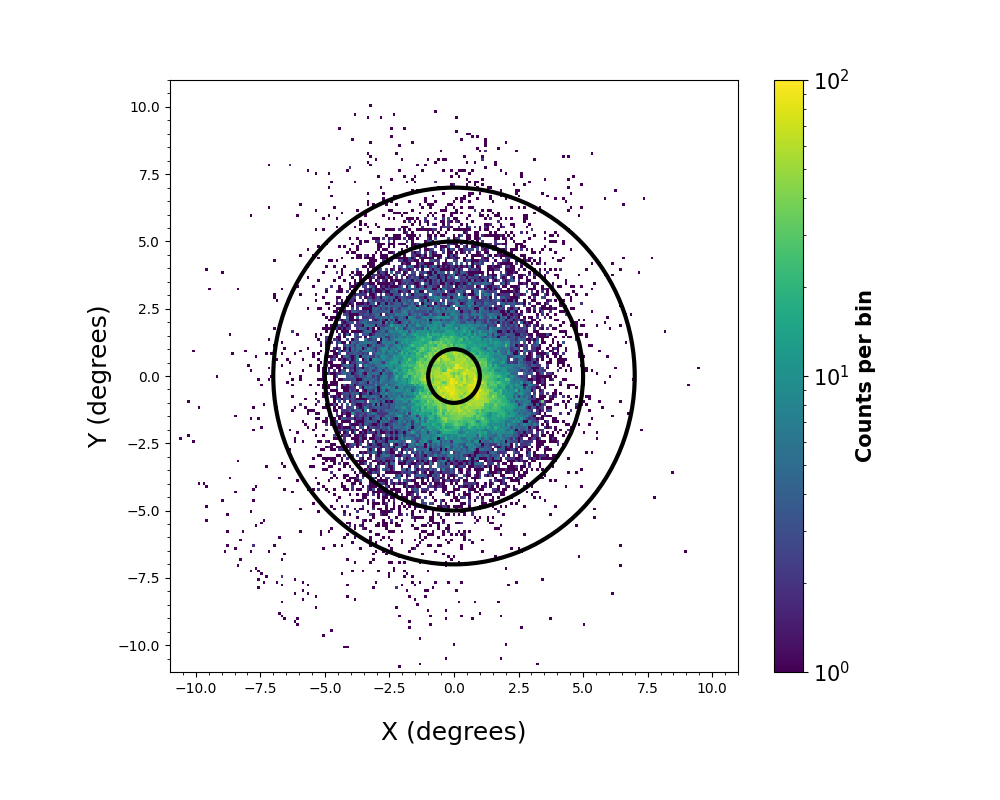}
      \includegraphics[width=1\columnwidth]{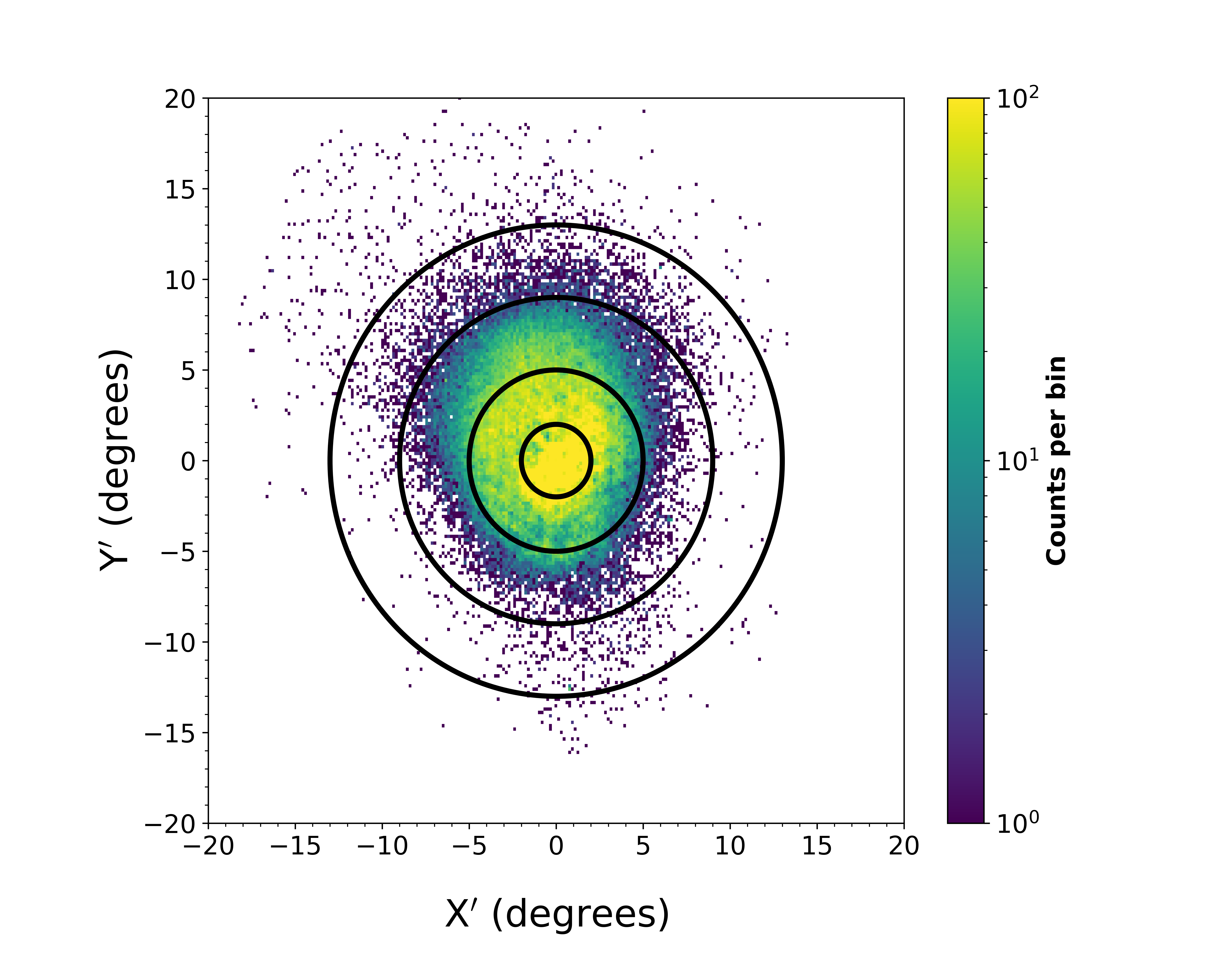}\\
      \includegraphics[width=1\columnwidth]{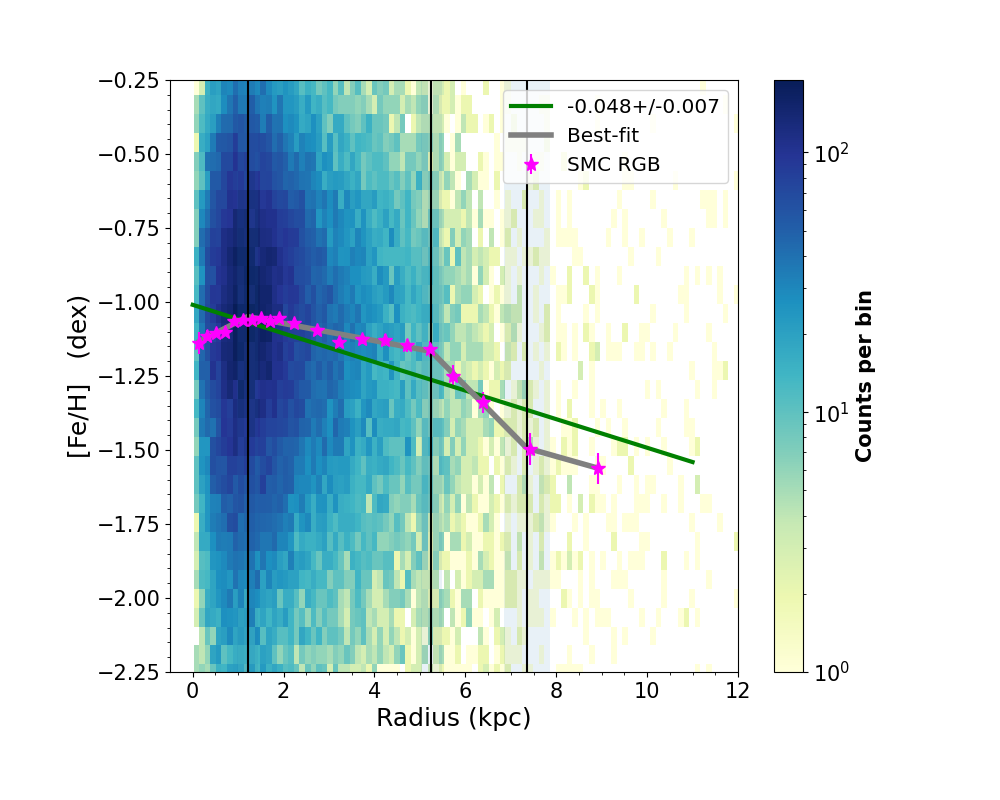}  
     \includegraphics[width=1\columnwidth]{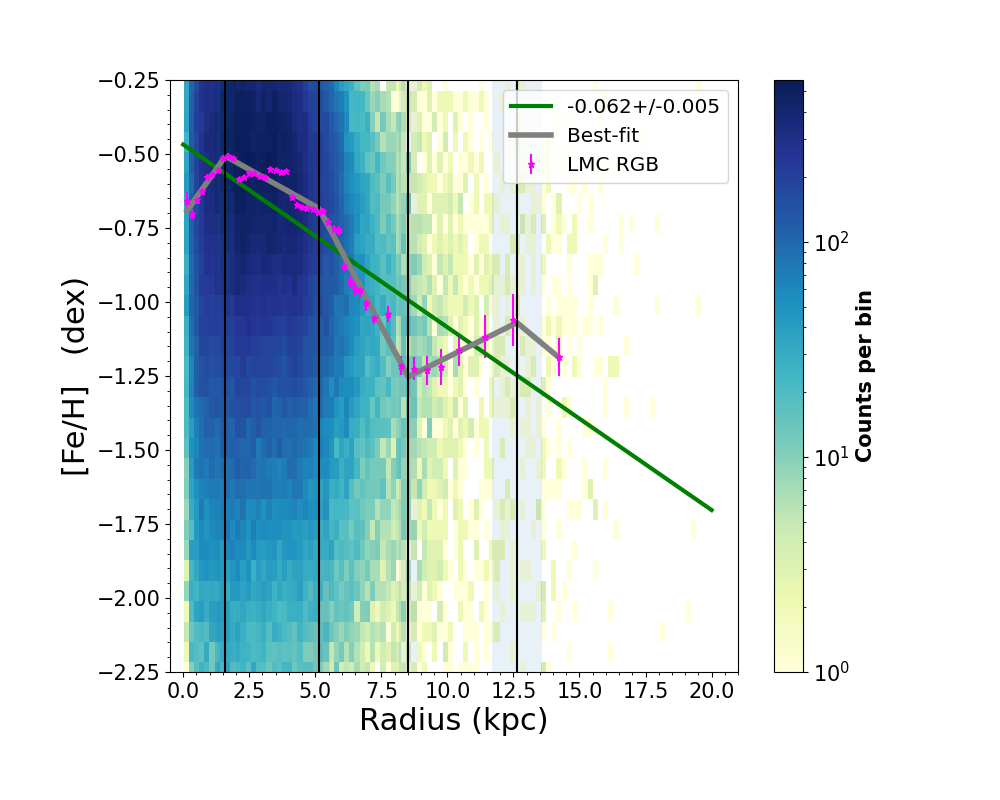} 
     \caption{Distribution of RGB stars within the SMC (top-left) and the LMC (top-right) with superimposed annular rings of breakpoints marked in the corresponding lower panels. The plots show the metallicity gradients for the SMC (bottom-left) and LMC RGB stars (bottom-right). The median metallicities of the annular bins are marked with pink stars along with their standard errors. Linear best-fits are marked with green and the piecewise-regression fits are marked with grey. The black vertical lines show the division of different segments or the location of the breaks in the gradient. In both panels, a Hess diagram shows the density distribution of the sources colour-coded such as blue showing the regions with the highest stellar density.}
    \label{fig:feh_rgb}
\end{figure*}
\label{rgbmap}

\subsection{Metallicity gradient of RGB stars}
\label{metallicty_gradient_rgb}

To further explore the metallicity distribution of the SMC and LMC RGB stars, we derive the metallicity gradients separately for the two galaxies as a function of distance from their corresponding centres. In the SMC, we first convert the distance from degrees into kpc by using 62 kpc $\times$ tan(radius), where 62 is the distance to the galaxy from \cite{degrijs2015smc}. Then, we radially bin the sources into 0.2 kpc steps up to 2 kpc from the centre. We use 0.5 kpc steps up to 6 kpc, 1 kpc steps up to 8 kpc and combine the rest of the sources until 11 kpc into a single sub-region. We choose radial bins of different radii due to the decreasing stellar density from the centre toward the outskirts. The number of sources in each bin ranges from around 550 to a few thousand in the central bins to around 270 in the outer ones. The metallicity gradient of the SMC RGB stars as a function of distance is shown in the bottom-left panel of Figure \ref{fig:feh_rgb}. For this, we calculate the median metallicities in each of the radial bins and also their respective standard errors. We use the Python module curve-fit to find the estimate of the best-fit metallicity gradient. We observe a linearly decreasing trend and a negative metallicity gradient of --0.048$\pm$0.007 dex kpc$^{-1}$. In the figure, the gradient is superimposed on the distribution of the estimated individual stellar metallicities. 

Although we can estimate the overall gradient by using a linear fit, it is evident from the median values that the slope changes are not uniform across the SMC but there are breaks in the gradient. To quantify these changes in the slope we use the Python module piecewise-regression or segmented regression \citep{piecewiseregression}. This module is particularly useful in estimating the different slopes in the dataset without having to fit different slopes in different regions. This module uses an iterative method to identify breakpoint positions and simultaneously fits a linear regression model to data that includes one or more breakpoints corresponding to the gradient's changes. This makes it computationally efficient and allows for robust statistical analysis. We iterate the fit over various numbers of breakpoints starting from zero which means no breaks in the overall gradient until the model converges. Then we obtain the Bayesian Information Criterion (BIC) which takes into account the value of the likelihood function to avoid over-fitting and choose the number of breaks where the BIC is minimum as the best-fit model. For the SMC, the best-fit model has three breaks and we have four regions with different gradient values. All these breakpoints are marked as black concentric circles in Figure \ref{fig:feh_rgb} (top-left). The different regions and their gradients are provided in Table \ref{tab:smc_pwlf_rkpc}. 

We followed the same procedure to estimate the metallicity gradient of the LMC RGB stars. One difference is that the LMC metallicity gradient is quoted as a function of the de-projected distance. Then using 50 kpc $\times$ tan(radius), where 50 is the distance to the galaxy from \cite{degrijs2014lmc} we obtain the de-projected distance in kpc. Also, in comparison with the SMC, our LMC sample is larger and hence we redefined the size of our radial bins as follows. We radially bin the sources into 0.2 kpc steps up to 7 kpc from the centre. Then, we use 0.5 kpc steps up to 10 kpc, 1 kpc steps up to 13 kpc and combine the rest of the sources until 20 kpc into a single sub-region. The median metallicity of each of these sub-regions is estimated along with their standard errors and shown in Figure \ref{fig:feh_rgb}. The linear best-fit for the overall metallicity gradient is also shown and suggests an increasing metallicity until $\sim$2.5 kpc from the centre. Although in these regions we can have the highest stellar density, \textit{Gaia} DR3 is incomplete and sources are still missing. The metallicity gradient corresponds to --0.062$\pm$0.005 dex kpc$^{-1}$. Our best-fit model for the segmented regression returns four breakpoints and hence five line segments with different gradients are as given in Table \ref{tab:lmc_pwlf_rkpc}. The overall metallicity gradient for the entire region is also provided at the bottom of the table. The breakpoints are marked as concentric circles in Figure \ref{fig:feh_rgb} (top-right). To facilitate the comparison with previous studies, we repeat the analysis by using radial bins in degrees. The resulting metallicity gradients for the SMC and LMC RGB stars are given in the top-left and top-right panels of Figure \ref{fig:feh_gradient}, respectively and the values are provided in the appendix.

\subsection{Metallicity gradient of supergiants stars}

\begin{figure*}
	\centering
	\includegraphics[width=1\columnwidth]{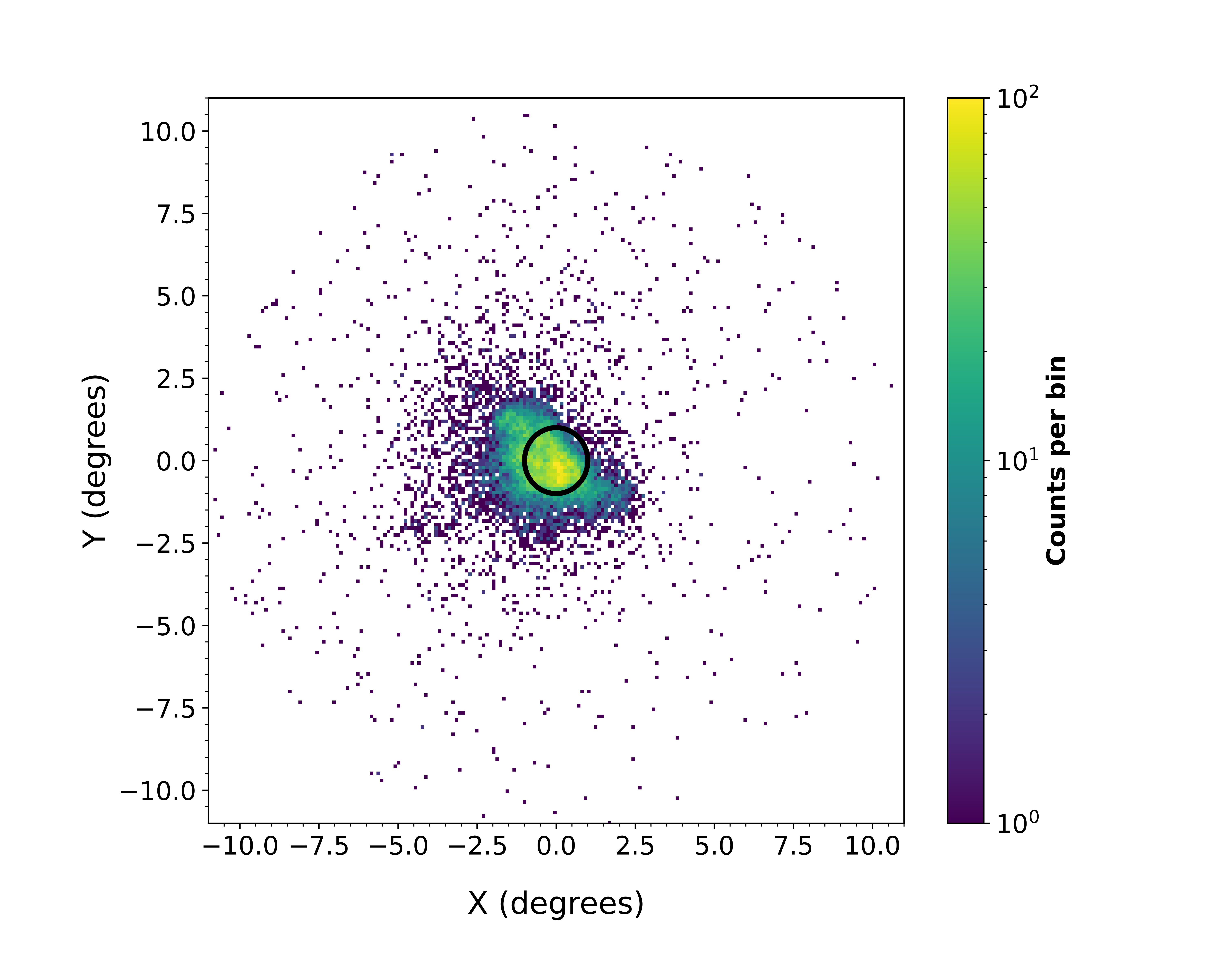}
	\includegraphics[width=1\columnwidth]{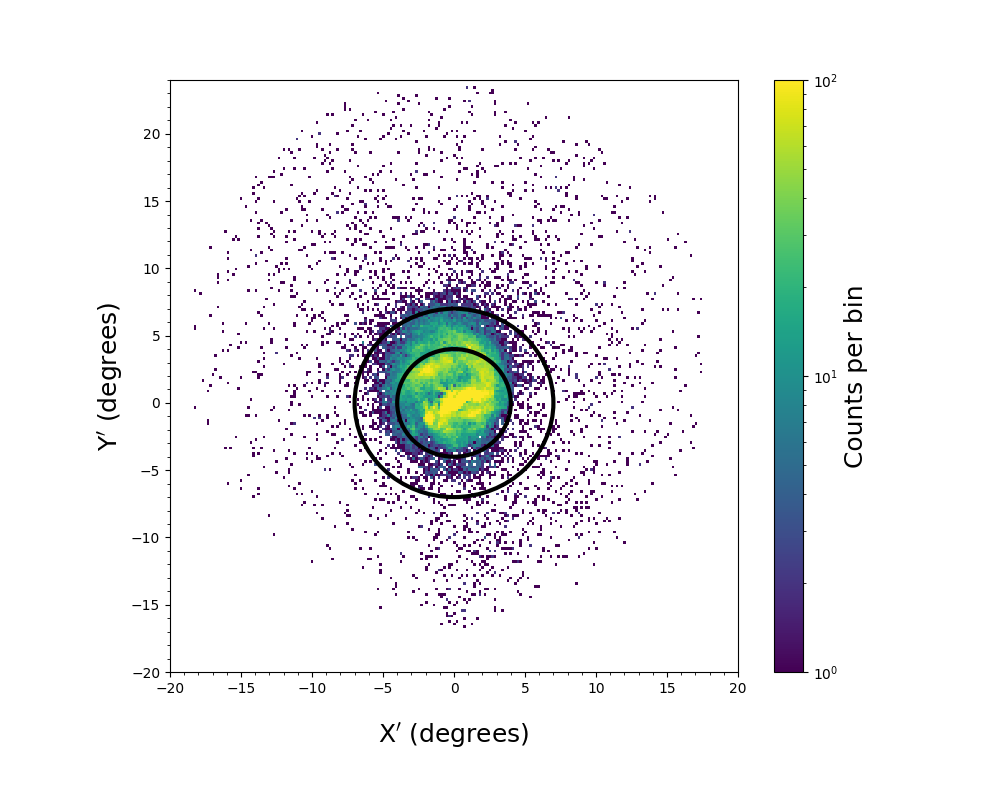}\\

	\includegraphics[width=1\columnwidth]{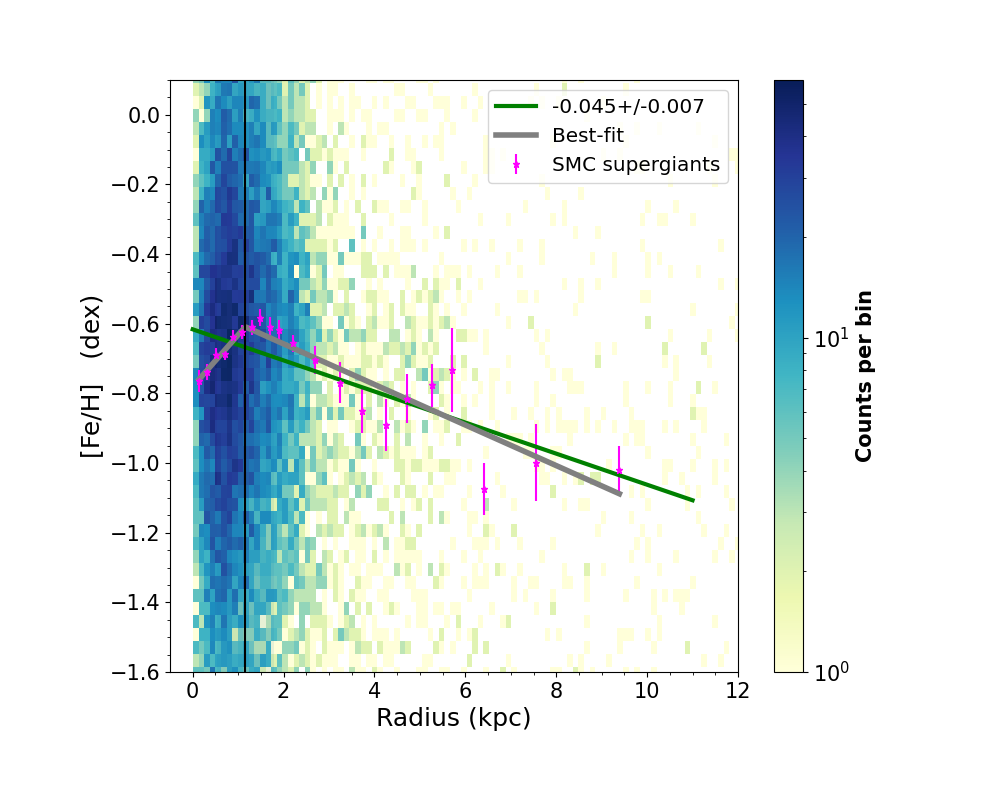}
	\includegraphics[width=1\columnwidth]{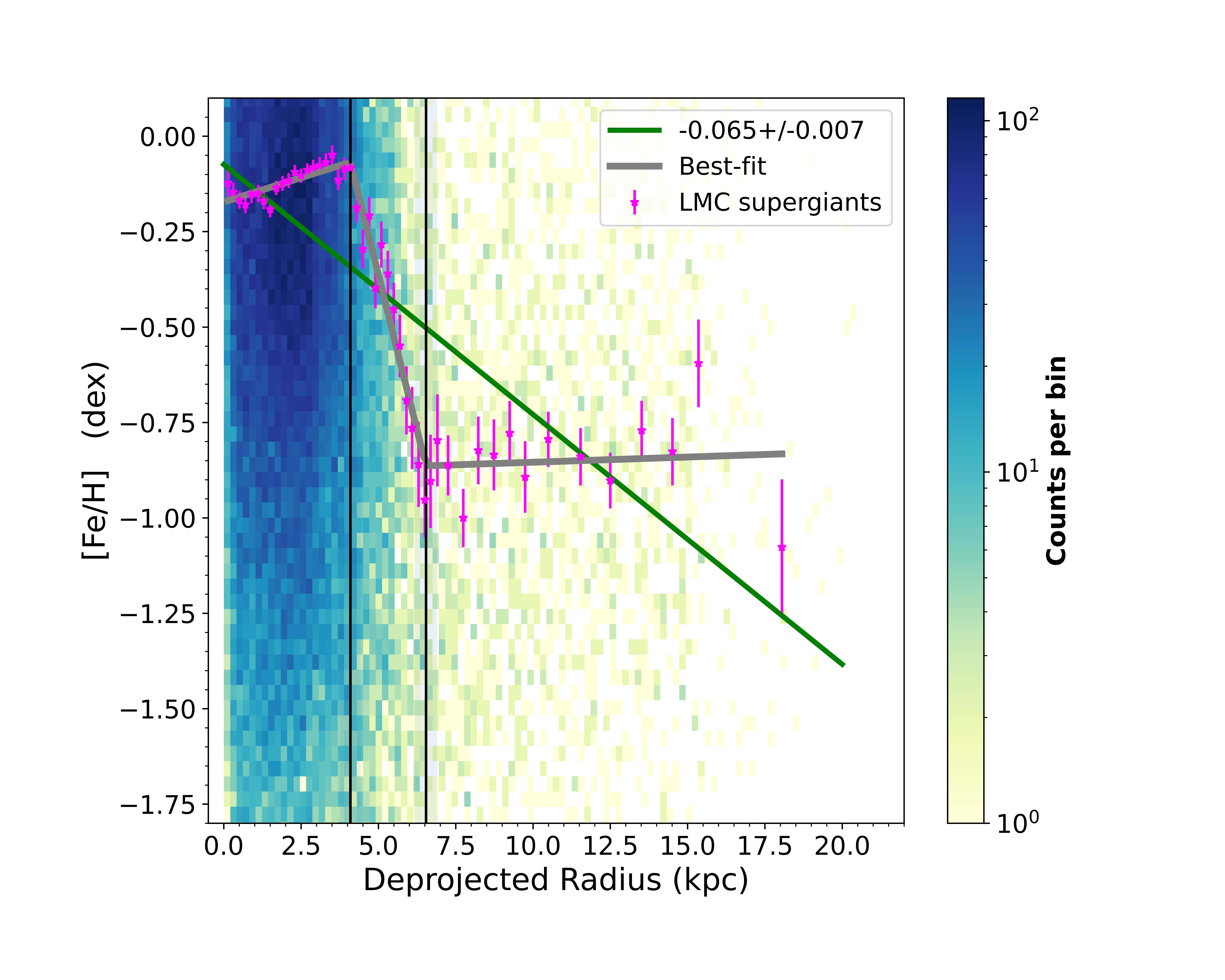}
	\caption{Same as Figure \ref{fig:feh_rgb} but for supergiants.}
    \label{fig:feh_supergiant}
\end{figure*}

We present here the metallicity gradient as a function of the distance (kpc) of the young supergiant stars in both the SMC and the LMC. We use the same radial binning that was applied to the SMC RGB sample. The metallicity gradients of the supergiant stars are shown in Figure \ref{fig:feh_supergiant} (bottom). We also use the same Python curve-fit module to fit the median metallicities of different radial bins and derive the best-fits. In the SMC, we find a linearly decreasing trend in the metallicity distribution and the estimated metallicity gradient is --0.045$\pm$0.007  dex kpc$^{-1}$. The best-fit for the piecewise-regression method has only one break around 1 kpc corresponding to two gradient values. We redefined our radial binning for the LMC supergiant sources. For the LMC supergiants, we radially bin the sources into 0.2 kpc steps up to 7 kpc from the centre. Then, we use 0.5 kpc steps up to 10 kpc, 1 kpc steps up to 16 kpc and combine the rest of the sources until 20 kpc into a single sub-region. In the LMC, we find a peculiar trend in the median metallicities of the supergiant stars which can possibly be due to their intrinsically non-uniform spatial distribution. Most of these sources are concentrated in the bar region and in the major spiral arm of the LMC. At about 5 deg from the LMC centre the stellar density drops rapidly where the median metallicities tend to decrease abruptly. Towards the outskirts of the LMC, the stellar density distribution is very low but more or less uniform and hence the median metallicity values seem to flatten. The estimated metallicity gradient from the LMC supergiants is --0.065$\pm$0.007 dex kpc$^{-1}$. The best-fit for the piecewise-regression is obtained with two breaks and three gradients. In the inner region, we have a positive metallicity gradient and then we have negative gradients with a break around 4 kpc. Beyond 7 kpc we obtain an almost flat gradient value. The different regions and their metallicity gradients for the SMC and the LMC supergiants are provided in Table \ref{tab:smc_pwlf_rkpc} and \ref{tab:lmc_pwlf_rkpc}, respectively. 

\begin{table}[]
	\centering
	\caption{Best-fit radial metallicity gradient for the SMC sources.}
	\label{tab:smc_pwlf_rkpc}
	\begin{tabular}{cr|cr}
		\hline \hline
		\multicolumn{2}{c}{RGB}                            & \multicolumn{2}{c}{Supergiants} \\ 
		Distance & Gradient & Distance & Gradient \\
		(kpc) & (dex kpc$^{-1}$) & (kpc) & (dex kpc$^{-1}$) \\
		\hline
		0 – 1            & 0.079$\pm$0.014   & 0 – 1            & 0.148$\pm$0.086   \\
		1 – 5            & --0.029$\pm$0.003  & R \textgreater 1 & --0.058$\pm$0.008  \\
		5 – 7            & --0.155$\pm$0.014  &                  &                   \\
		R \textgreater 7 & --0.043$\pm$0.011  &                  &                   \\
		0 – 11           & --0.048$\pm$0.007 & 0 – 11           & --0.045$\pm$0.007 \\
		\hline
	\end{tabular}
\end{table}

\begin{table}[]
	\centering
	\caption{Best-fit radial metallicity gradient for the LMC sources.}
	\label{tab:lmc_pwlf_rkpc}
	\begin{tabular}{cr|cr}
		\hline \hline
		\multicolumn{2}{c}{RGB}                           & \multicolumn{2}{c}{Supergiants} \\ 
		Distance & Gradient & Distance & Gradient \\
		(kpc) & (dex kpc$^{-1}$) & (kpc) & (dex kpc$^{-1}$) \\
		\hline
		0 – 2             & 0.127$\pm$0.026   & 1 – 4            & 0.025$\pm$0.014   \\
		2 – 5             & --0.050$\pm$0.008  & 4 – 7            & --0.325$\pm$0.036  \\
		5 – 9             & --0.169$\pm$0.011 & R \textgreater 7 & 0.003$\pm$0.006   \\
		9 – 13            & 0.044$\pm$0.016   &                  &                   \\
		R \textgreater 13 & --0.074$\pm$0.027  &                  &                   \\
		0 -- 20           & --0.062$\pm$0.005 & 0 – 20           & --0.065$\pm$0.007 \\
		\hline
	\end{tabular}
\end{table}

\subsection{Metallicity catalogue}

The final catalogues for all the sources with estimated [Fe/H], recalibrated using the APOGEE values, are published with this study.  As an example, the estimated [Fe/H] values along with the reddening values and other parameters characterising the individual sources (Gaia DR3 Source IDs, coordinates at the epoch J2016.0, Str\"{o}mgren magnitudes, and distances from the centre of the galaxies) are shown in Table \ref{tab:feh}.

\begin{table*}[h]
\centering
\setlength{\tabcolsep}{2.5pt}
\caption{Catalogue of the estimated {[}Fe/H{]} for RGB stars in the SMC.}
\label{tab:feh} 
\begin{tabular}{cccccccccccc}
\hline \hline
Gaia Source ID &
\begin{tabular}[c]{@{}c@{}}RA\tablefootmark{a} \\ (deg)\end{tabular} &
\begin{tabular}[c]{@{}c@{}}Dec \\ (deg)\end{tabular} &
\begin{tabular}[c]{@{}c@{}}v\\  (mag)\end{tabular} &
\begin{tabular}[c]{@{}c@{}}b \\ (mag)\end{tabular} &
\begin{tabular}[c]{@{}c@{}}y\\  (mag)\end{tabular} &
\begin{tabular}[c]{@{}c@{}}D \\ (deg)\end{tabular} &
\begin{tabular}[c]{@{}c@{}}D\\ (kpc)\end{tabular} &
\begin{tabular}[c]{@{}c@{}}E(B--V)\\  (mag)\end{tabular} &
\begin{tabular}[c]{@{}c@{}}A$\_V$\\  (mag)\end{tabular} &
\begin{tabular}[c]{@{}c@{}}{[}Fe/H{]}\\ (dex)\end{tabular} &
\begin{tabular}[c]{@{}c@{}}{[}Fe/H{]}$\_err$\\ (dex)\end{tabular} \\
\hline
4688748623761036544 & 5.05237 & --73.28698 & 18.829 & 17.607 & 16.816 & 2.413 & 2.612 & 0.034 & 0.106 & --0.851 & 0.418 \\
4688746978787018496 & 4.79228 & --73.23919 & 19.181 & 17.931 & 17.065 & 2.482 & 2.687 & 0.029 & 0.088 & --1.229 & 0.397 \\
4688748902932208256 & 5.01639 & --73.26982 & 19.341 & 18.045 & 17.223 & 2.421 & 2.621 & 0.034 & 0.106 & --0.811 & 0.502 \\
4688748078298643072 & 4.75311 & --73.19900 & 19.238 & 18.081 & 17.320 & 2.490 & 2.696 & 0.047 & 0.144 & --0.829 & 0.549 \\
4688749006011415424 & 5.11183 & --73.25140 & 19.395 & 18.305 & 17.702 & 2.391 & 2.589 & 0.034 & 0.106 & 0.147  & 0.730\\ 
\hline
\end{tabular}
\tablefoot{
The full table and a similar table for the supergiants as well as tables for the LMC sources are available at the CDS.\\
\tablefoottext{a}{The RA and Dec coordinates are truncated here for the purpose of visualization and are given as obtained from \textit{Gaia} DR3 in the online version.}}

\end{table*}

\begin{figure*}
	\centering
	\includegraphics[scale=0.3]{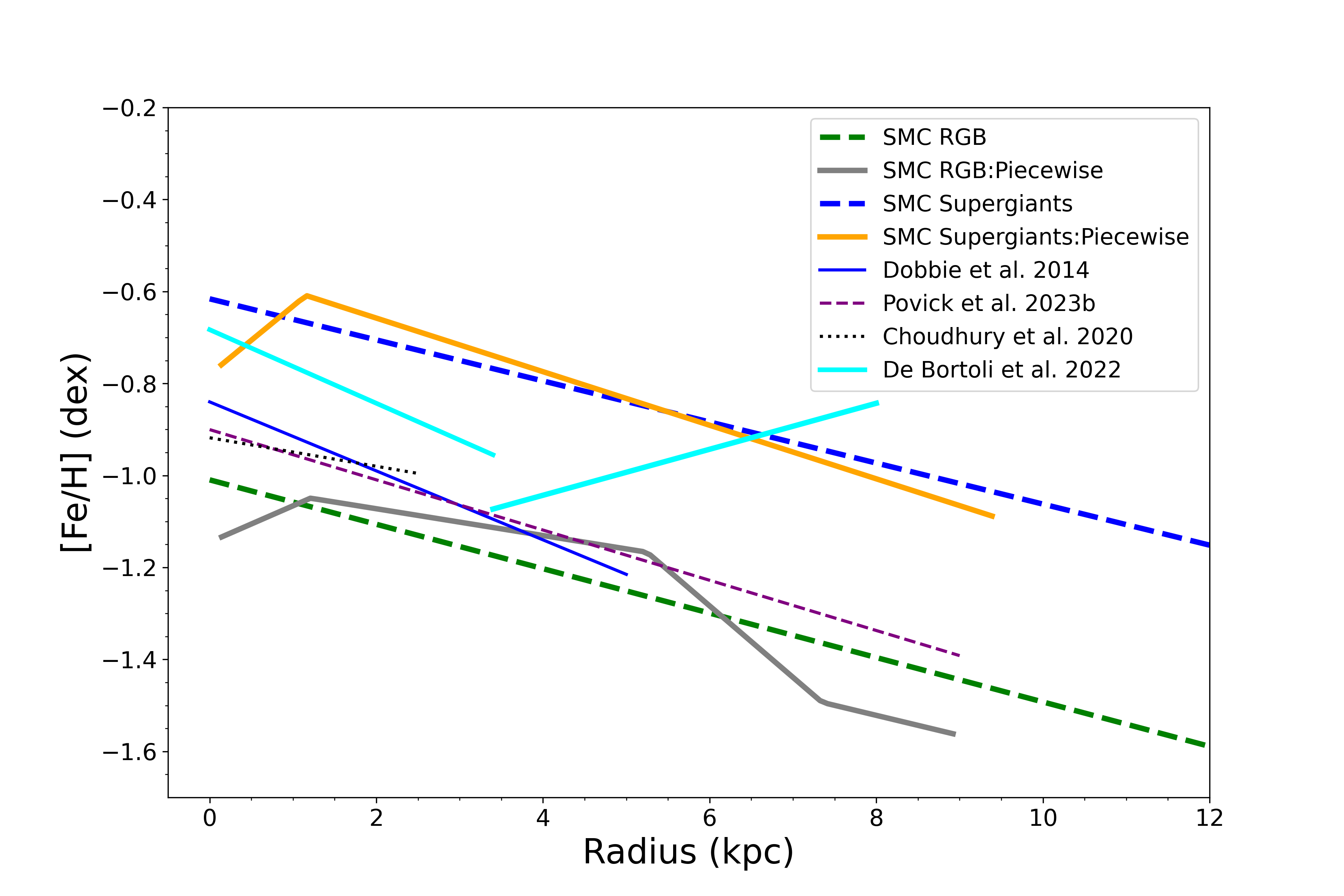}	
	\includegraphics[scale=0.3]{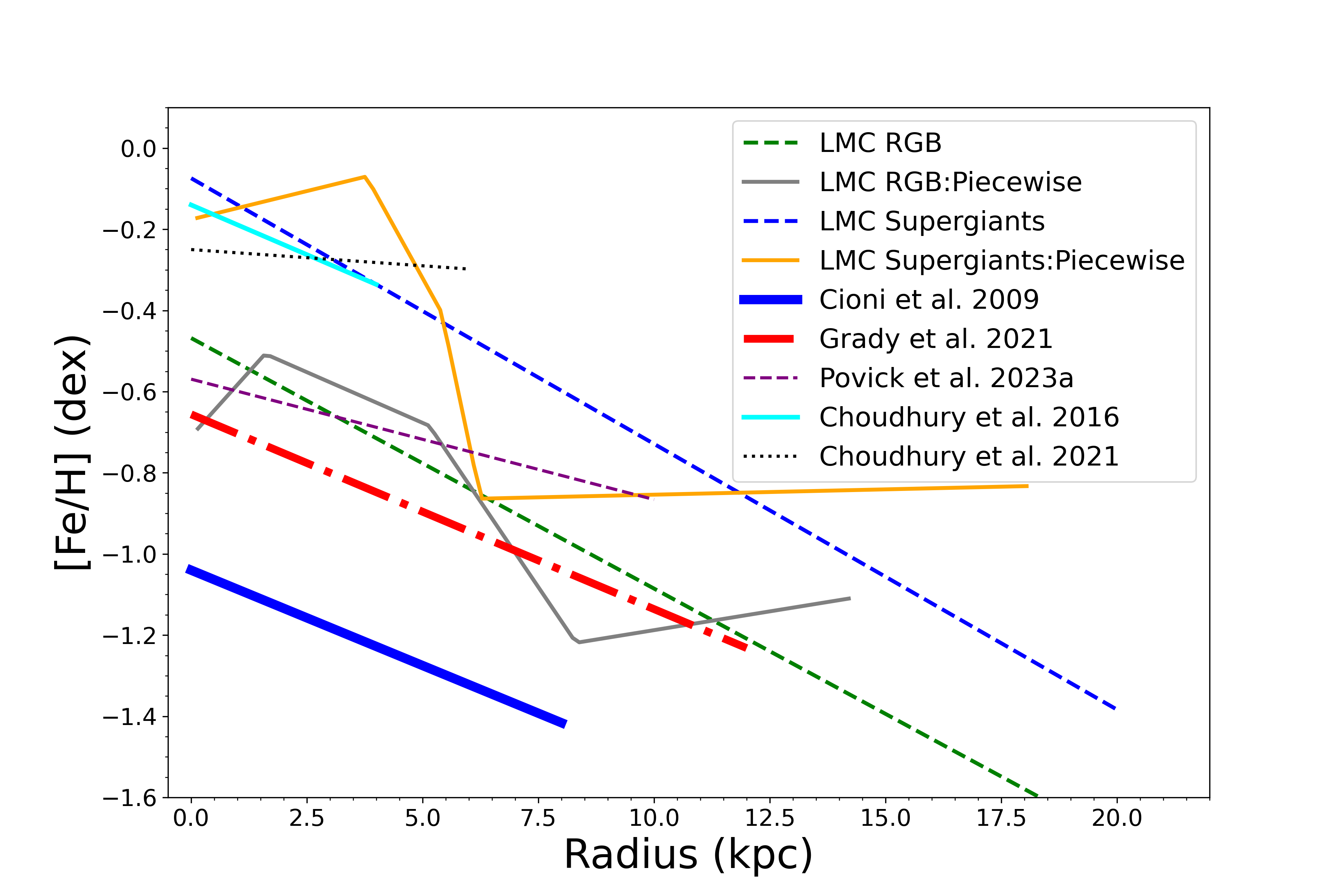}
	\caption{Comparison of the radial metallicity gradients of the SMC (left) and the LMC (right) with the literature is shown here.}
	\label{fig:feh_gradient_comparison}
\end{figure*}

\section{Discussion and Conclusions}
\label{summary}
In this work, we estimate the photometric metallicities from \textit{Gaia} DR3--XP spectra using the synthetic Str\"{o}mgren photometry. We use the different calibration relations from the literature to derive the [Fe/H] abundance of the RGB (old, >3 Gyr old) and the supergiant (young, <300 Myr old) stars in the SMC and the LMC. We also compare our photometric metallicity estimates with the spectroscopic metallicities from APOGEE to validate our method. We use our photometric metallicities to create metallicity maps with a spatial resolution of 0.25--0.5 deg$^2$ covering a wide area of the Clouds for both stellar populations. We also compute metallicity gradients with respect to the centres within both galaxies. Our choice of the centres should not influence the gradients because the central and smallest bins correspond to a circular diameter of 0.4 kpc (about 0.4 deg), which includes other estimated centres (e.g. \citealp{niederhofer2021_smc, niederhofer2022_lmc}). In the left panel of Figure \ref{fig:feh_gradient_comparison}, we show the comparison of the radial metallicity gradients obtained in our study for the SMC. 

\subsection{SMC gradient}
In our study, the estimated overall metallicity gradient of the SMC RGB stars is --0.048$\pm$0.007 dex kpc$^{-1}$, which is very similar to the gradient obtained for the supergiants (--0.045$\pm$0.007 dex kpc$^{-1}$) out to 11 deg ($\sim$10 kpc) from the SMC centre. \cite{povick2023smc} explored the metallicity estimates of the RGB stars from APOGEE and found a similar overall metallicity gradient of --0.0545$\pm$0.0043 dex kpc$^{-1}$ for the SMC stars out to $\sim$9 kpc. Previously \cite{choudhury2020smcvmc} used the RGB slope as an indicator of the average metallicity and estimated a metallicity gradient of --0.031$\pm$0.005 dex deg$^{-1}$ within 2.5 deg of the centre of the galaxy finding also a rather flat gradient beyond this distance. This is shallower than our current overall estimates but is in good agreement with the gradient we find from RGB stars located within 1--5 deg --0.029$\pm$0.003 dex kpc$^{-1}$. The steeper gradient we find up to 1 kpc from the centre (0.079$\pm$0.014 dex kpc$^{-1}$) might be influenced by the incompleteness of our dataset due to crowding. In the outer regions, the gradient first decreases steeply, the slope is higher than in the innermost regions, and then show a flattening. A previous study by \cite{Parisi2016} also found, using star clusters, a change of slope at about 5 kpc suggesting an upturn of the gradient until 8 kpc. In this less-gravitationally bound region, the imprints of dynamical interactions are strong and there are existing sub-structures (e.g.: \citealp{El_Youssoufi_2021}) which may influence the metallicity distribution. The supergiants also show an increase and then a decrease in the metallicity gradient. Our estimated best-fit for the inner region up to 1 kpc is 0.148$\pm$0.086 dex kpc$^{-1}$ and beyond that --0.058$\pm$0.008 dex kpc$^{-1}$. The only breakpoint we find in the centre could again be the result of crowding and not an actual breakpoint in the metallicity gradient. 

The spectroscopic study by \cite{dobbiefeh} estimated an RGB metallicity gradient of --0.075$\pm$0.011 dex deg$^{-1}$ out to 5 deg from the centre.  Later, \cite{taibi2022} derived the radial metallicity gradients of several Local Group galaxies recalibrating the metallicity estimates from \cite{dobbiefeh}. Although they have not provided metallicity gradients for different regions across the SMC, their smoothed SMC radial metallicity profile excellently matches the metallicity gradient of SMC RGB stars in our study. Interestingly, they also see a positive gradient up until 1 kpc from the SMC centre, which was not clearly seen or noted in other studies. Then, their gradient decreases until 4 kpc, slight increases until 5 kpc and then decreases further until 7 kpc in line with the trends found in this study. Note that a change of slope at $\sim$4 kpc could be deduced from Figure \ref{fig:feh_rgb} but it is not sufficiently strong to be picked up in our analysis as a separate breakpoint. \cite{debortoli2022} estimated a negative gradient of --0.08$\pm$0.03 dex deg$^{-1}$ in the inner region ($<$ 3.4 deg), but a positive or null gradient in the outer region 0.05$\pm$0.02 dex deg$^{-1}$ derived from the analysis of SMC field stars. Their values are comparable with the main-body cluster metallicity gradients derived in the same study. However, they do not find any separate metallicity gradient for their metal-rich and metal-poor clusters.

Although our estimated overall metallicity gradient seems to overlap with previous studies, it can still be seen that the gradient in different regions differ considerably. One reason could be how we are projecting the SMC stars and estimating the gradients, as the SMC has a large line-of-sight depth and a complex structure. Taking this into account, \cite{dias2022} showed that any projected radial metallicity gradient cannot be seen when using the real three-dimensional (3D) physical distances to the clusters. Our dataset do not consist of any standard candles to estimate distances based on them. Also at the distance of the Clouds, estimating the 3D distances for individual stars is not trivial.

\subsection{LMC gradient}
For the RGB stars across the entire LMC, we obtain an overall metallicity gradient of --0.062$\pm$0.005 dex kpc$^{-1}$, whereas for supergiants we obtain a negative gradient of --0.065$\pm$0.007 dex kpc$^{-1}$. In the right panel of Figure \ref{fig:feh_gradient_comparison}, we show the comparison of the radial metallicity gradients obtained in our study for the LMC. A recent study by \cite{povick2023lmc} estimated a metallicity gradient of the LMC to be --0.02966$\pm$0.00171 dex kpc$^{-1}$ out to a radial distance of $\sim$10 deg using spectra of RGB stars from APOGEE, which is shallower than our estimations. They also detect a steepening of the abundance gradients in the inner region of the galaxy for populations younger than 2 Gyr. \cite{grady2021} utilised the \textit{Gaia} DR2 data and estimated metallicities using a machine learning method to obtain --0.048$\pm$0.001 dex kpc$^{-1}$ out to $\sim$12 deg.  Also, a previous estimate from \cite{cioni2009} using the AGB stars in the LMC found a linearly decreasing gradient of --0.047$\pm$0.003 dex kpc$^{-1}$ out to $\sim$8 kpc from the centre. A study by \cite{choudhury2015} estimated photometric metallicities and found a similar [Fe/H] trend in the LMC as --0.049$\pm$0.002 dex kpc$^{-1}$ out to $\sim$4 kpc. In a later study using the VMC data of the RGB stars, \cite{choudhury2021lmc} estimated a metallicity gradient of --0.008$\pm$0.001 dex kpc$^{-1}$ out to a radius of $\sim$6 kpc from the LMC centre. This value compares well with our measurement in 2--5 kpc whereas the overall gradient is somewhat lower compared to the previous studies probably because our study extends to 20 deg and includes the outskirts of the galaxy where there is a relatively large number of metal-poor stars.\\ 
\indent Our best-fit for the piecewise-regression fit estimates of the metallicity of RGB stars results in four different gradients for the entire LMC. In the very centre until 2 kpc, we get a positive gradient of 0.127$\pm$0.026 dex kpc$^{-1}$. This is the region that mostly encompasses the bar, which has both young and old stars. Also, this is the region that is subjected to crowding. In a way the first breakpoint in the gradient more or less separates the bar from the rest of the galaxy. From 2 to 5 kpc, representing the inner disc of the LMC, we find a gradient of  --0.050$\pm$0.008 dex kpc$^{-1}$. A recent study by \cite{Rathore_2025} finds that the angular momentum section of the inner disc is misaligned with respect to the outer disc which can be reproduced in simulation by a recent collision with the SMC. In our study, a break in the metallicity gradient at 5 kpc corresponds to a transition in the amount of misalignment which suggests a possible link between the two behaviours. The misaligned inner disc, influenced by the SMC interaction, may have played a role in the chemical enrichment processes, leading to the differing metallicity profiles. In the regions from 5 to 9 kpc, we find a very steep decrease in the gradient corresponding to --0.169$\pm$0.011 dex kpc$^{-1}$ and in the outskirts beyond 9 until 13 kpc, we obtain a positive gradient of 0.044$\pm$0.016 dex kpc$^{-1}$. This outer disc region hosts different tidal sub-structures and has been mostly influenced by the interactions from the MW and also the SMC, its closest companion. Beyond 13 kpc, we again see a negative gradient of --0.074$\pm$0.027 dex kpc$^{-1}$.\\ 
\indent Our best-fit piecewise-regression fit for the supergiants provides two breaks in the metallicity gradient and the supergiants looks more complex than the RGB stars. In the inner region up to 4 kpc, we obtain a positive gradient of 0.025$\pm$0.014 dex kpc$^{-1}$. Similarly to the RGB stars, we can see that the bar region is enclosed within the first breakpoint. For the region between 4 to 7 kpc, we see a much steeper gradient of --0.325$\pm$0.036 dex kpc$^{-1}$. This region includes part of the prominent northern spiral arm, where young stars are forming. Moreover, it exhibits a significant decline in stellar density, which leads to the steep gradient observed here. Additionally, the breakpoint around 4 kpc also suggests a link with the misalignment of the angular momentum \citep{rathore2025}, which implies a connection between the LMC's dynamical response to the collision and its chemical evolution. Beyond 7 kpc, we do not have many stars, but the gradient seems to flatten into 0.003$\pm$0.006 dex kpc$^{-1}$.\\
\indent \cite{feast2010} derived the metallicity gradient out to 6 kpc from the LMC centre using two different period-metallicity relations for RR Lyrae stars. They estimated a linearly decreasing gradient of --0.0104$\pm$0.0021 dex kpc$^{-1}$ and --0.0145$\pm$0.0029 dex kpc$^{-1}$, whereas \cite{haschke2012rrlyrae} obtained a gradient of --0.03$\pm$0.07 dex kpc$^{-1}$ using RR Lyrae stars up to 14 kpc from the LMC centre. They also estimated the gradient only in the innermost 8 kpc and obtained a shallower value of --0.010$\pm$ 0.014 dex kpc$^{-1}$. \cite{wagnerkaiser2013} derived instead a  gradient of --0.0270$\pm$0.02 dex kpc$^{-1}$. In the inner 2--5 kpc we also find shallower gradient, from the RGB stars, compared to the measurements by \cite{feast2010} and \cite{haschke2012rrlyrae}. By including regions further away we also find a decreasing trend for the overall distribution of stars. We note that the RR Lyrae stars are at least 10 Gyr old and represent a stellar population that is on average older than our RGB stars.

\subsection{Concluding remarks}

The main goal of our study is to validate the potential usage of \textit{Gaia} DR3--XP spectra to estimate the individual [Fe/H] values across a wide area encompassing the Clouds. This method can be applied to sources fainter then 17.65 mag from the next \textit{Gaia} data releases to study even larger samples and confirm or refine our findings. The metallicity gradients estimated in this work and in comparison with the literature values show that different tracers do not show the same trends. Compared to previous studies, different trends are also present for tracers that cover different spatial regions. The metallicity gradient results from the combination of different physical processes: the chemical enrichment from stellar evolution, the accretion of gas or stars and the process of radial migration as well as dynamical interaction displacing stars within galaxies. The present-day LMC and SMC have a complex structure and to quantify which process dominates at a given distance and for a given stellar populations we need to also study azimuthal variations, which we plan to address in our next study focused on sub-structures.

In the near future, the [Fe/H] as well as the abundance of other chemical elements for a large number of stars making up the range of stellar populations present in the Clouds, will be obtained with the new multi-object spectroscopic facilities 4MOST \citep{dejong2019} and MOONS \citep{gonzalez2020}. In particular, 4MOST will target about 200\,000 RGB stars and 60\,000 supergiants across $\sim$1000 deg$^2$ from the One Thousand and One Magellanic Clouds fields survey \citep{cioni2019}, whereas MOONS will focus on RGB stars at specific locations during guaranteed-time observations. These spectra combined with photometric investigations will provide an unprecedented view of the system increasing not only the size of the statistical spectroscopic samples, but also the number of diagnostics to study its origin and evolution. \cite{lu2023} has shown the possibility of inferring birth radii for individual stars using cosmological simulations in galaxies like the LMC. They obtain $\sim$25$\%$ median uncertainties for individual stars if accurate metallicities and ages are available further supporting a promising application of our results. Our findings highlight the complexity of metallicity gradients and their dependence on different stellar tracers, spatial coverage, and physical processes shaping the evolution of the Clouds.

\begin{acknowledgements}
We thank the referee for their constructive and insightful comments on our paper. AOO acknowledges support from the Indian Institute of Astrophysics during her visit. SS acknowledges support from the Science and Engineering Research Board of India through a Ramanujan Fellowship and support from the Alexander von Humboldt Foundation. AOO is grateful to the European Space Agency (ESA) for support via the Archival Research Visitor Programme during her stay at ESTEC. BD acknowledges support by ANID-FONDECYT iniciación grant No. 11221366 and from the ANID Basal project FB210003. This work has made use of data from the ESA mission
{\it Gaia} (\url{https://www.cosmos.esa.int/gaia}), processed by the {\it Gaia}
Data Processing and Analysis Consortium (DPAC,
\url{https://www.cosmos.esa.int/web/gaia/dpac/consortium}). Funding for the DPAC
has been provided by national institutions, in particular, the institutions
participating in the {\it Gaia} Multilateral Agreement.
\end{acknowledgements}

\bibliographystyle{aa} 
\bibliography{reference} 

\begin{appendix}

\section{Metallicity recalibration using APOGEE}
\label{apogee}
\begin{figure}[htb]
	\centering
	\includegraphics[width=0.7\columnwidth]{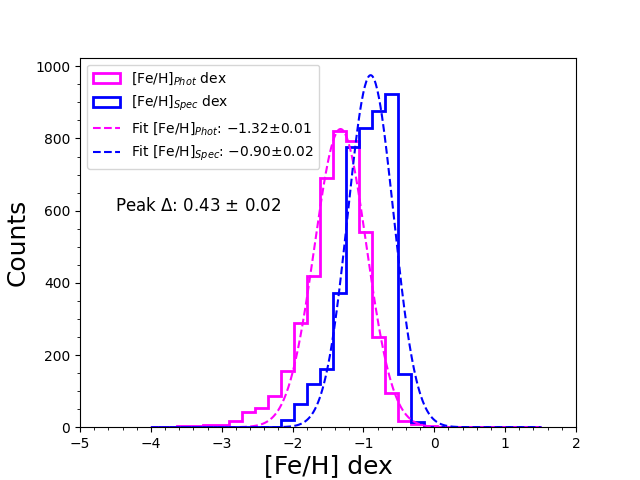}\\
	\includegraphics[width=0.7\columnwidth]{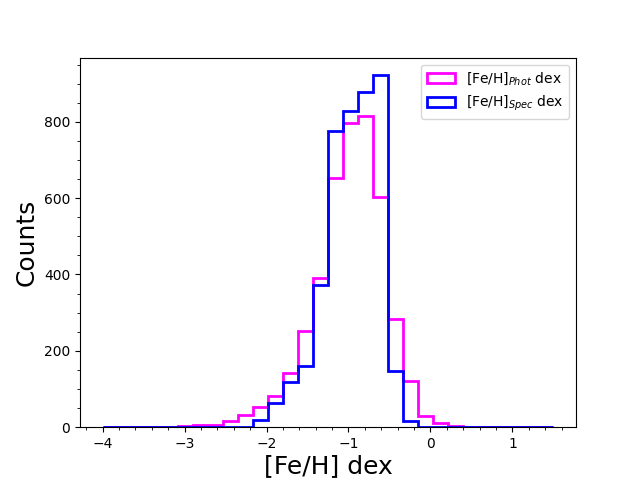}  
	\caption{Histograms of the LMC and SMC sources used for recalibration. On the top panel, the estimated photometric (Phot; blue) metallicities are plotted together with the spectroscopic (Spec; pink) metallicities as measured. On the bottom panel photometric metallicities are exploited after the addition of systematic difference.}
	\label{fig:rgb_hist_recalib}
\end{figure}

\begin{figure}[htb]
	\centering
	\includegraphics[width=0.7\columnwidth]{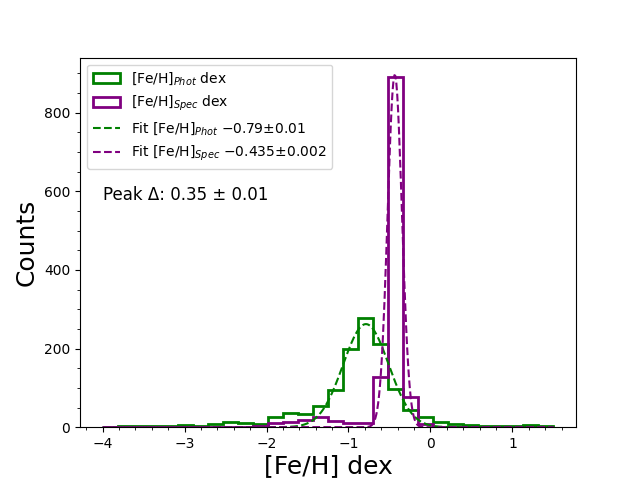}\\
	\includegraphics[width=0.7\columnwidth]{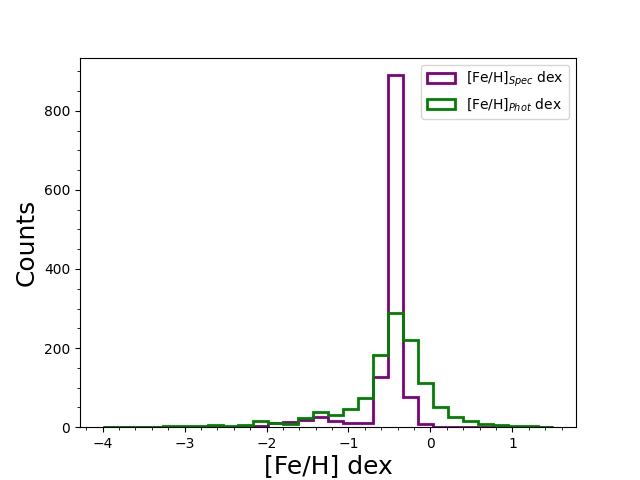}  
	\caption{Same as Figure \ref{fig:rgb_hist_recalib} but for the supergiant sources.}
	\label{fig:bl_hist_recalib}
\end{figure}

\begin{figure}[htb]
	\centering 
	\includegraphics[width=0.65\columnwidth]{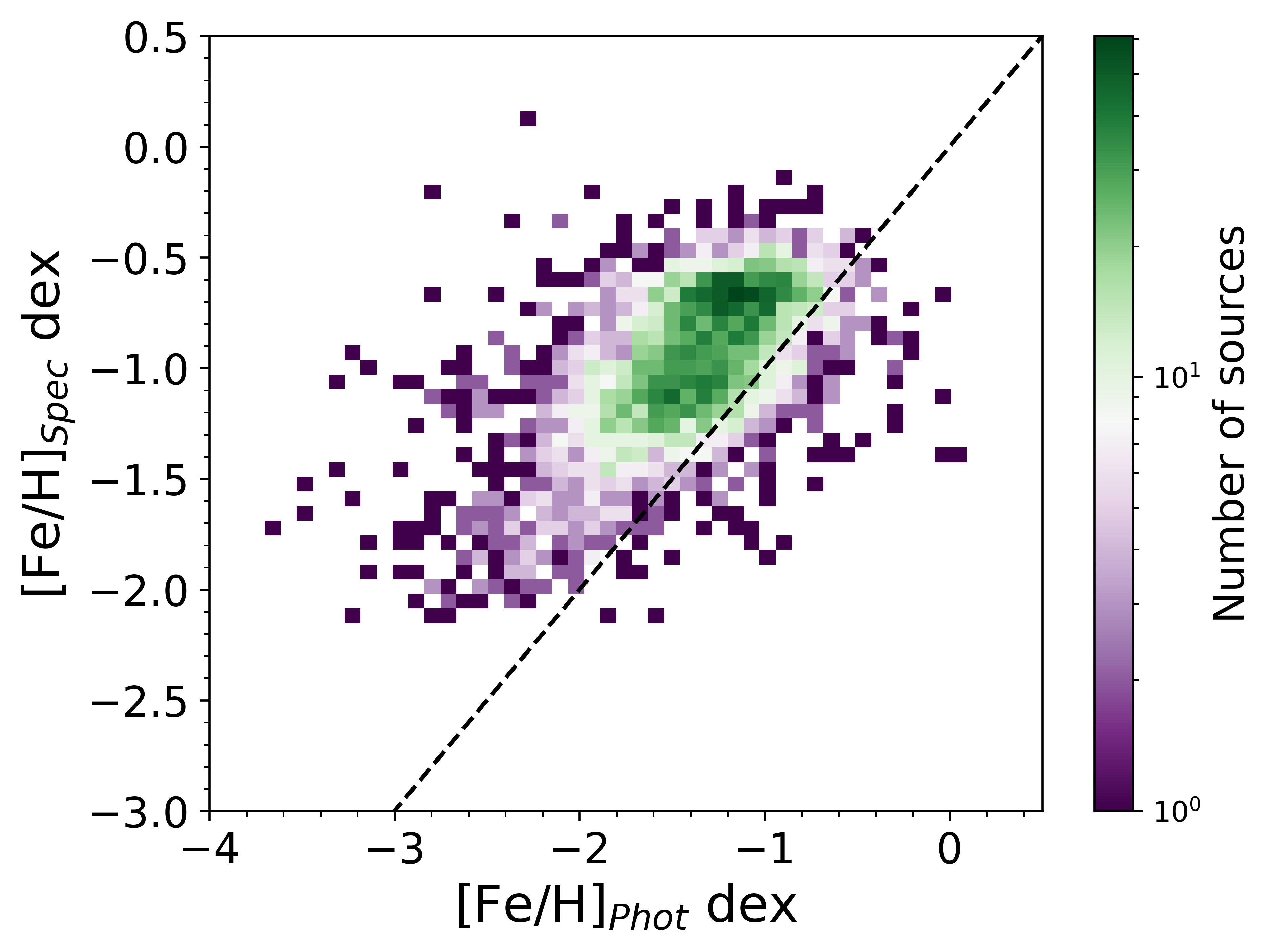}\\	
	\includegraphics[width=0.65\columnwidth]{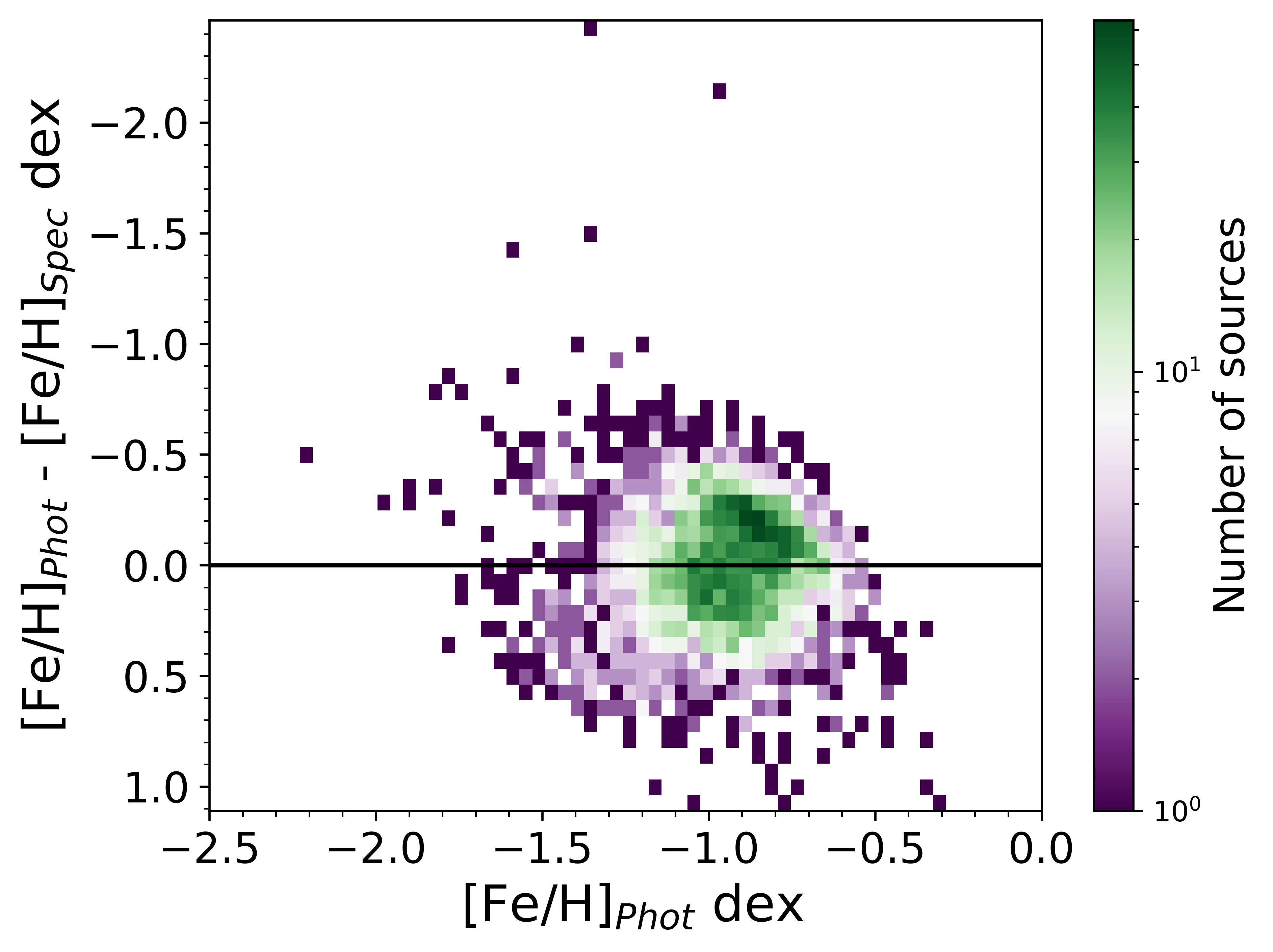}
	\caption{Photometric [Fe/H] vs spectroscopic [Fe/H] of the RGB sources are shown on the top panel. The dotted line marks the one-to-one relation. In the bottom panel, photometric [Fe/H] (recalibrated with the slope and constant) is plotted against the difference between them. The colour bar from violet to green shows the increase in stellar density.}
	\label{fig:rgb_recalib_hess}
\end{figure}

\begin{figure}[htb]
	\centering 
	\includegraphics[width=0.65\columnwidth]{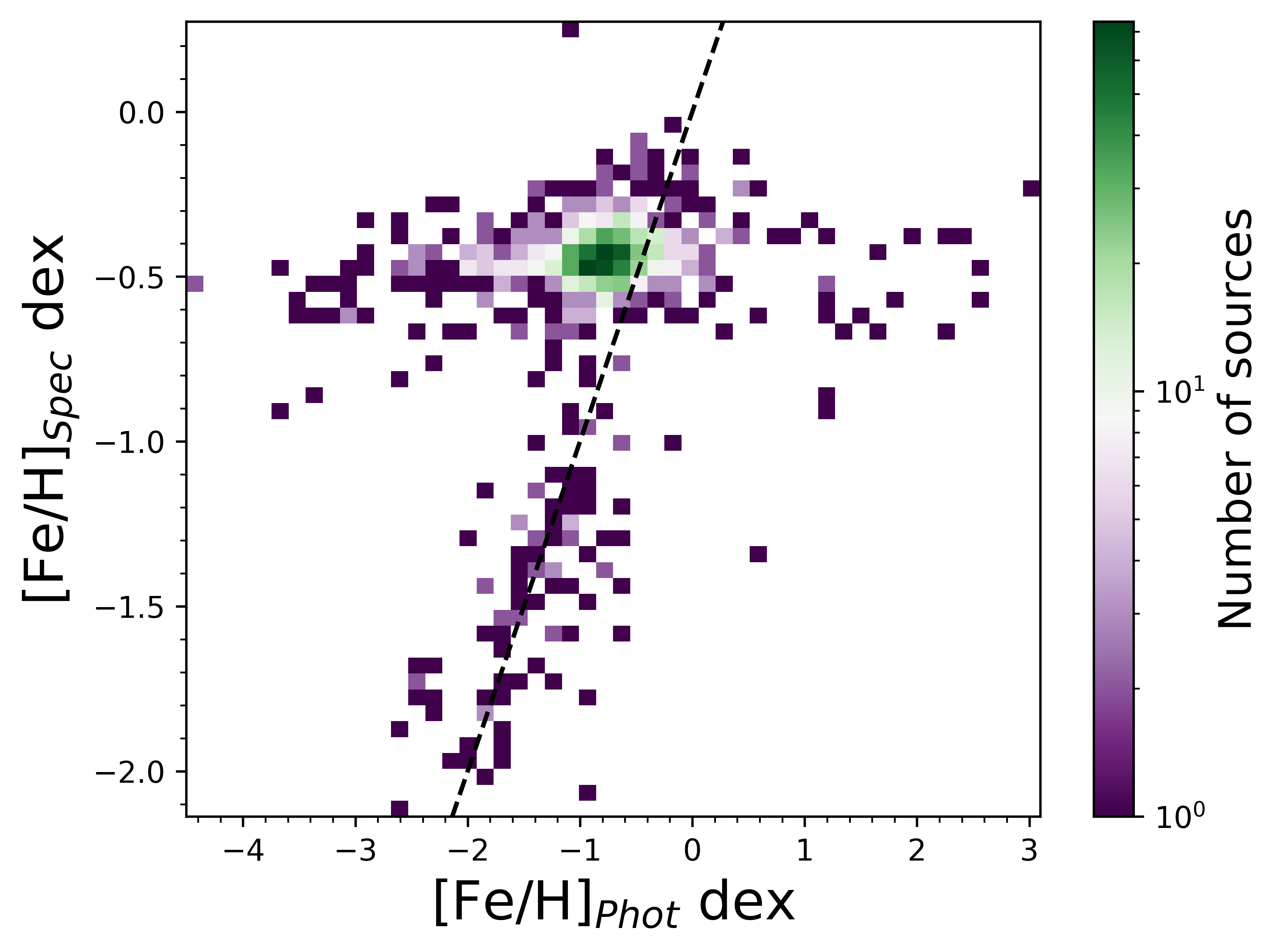}\\     	 	 \includegraphics[width=0.68\columnwidth]{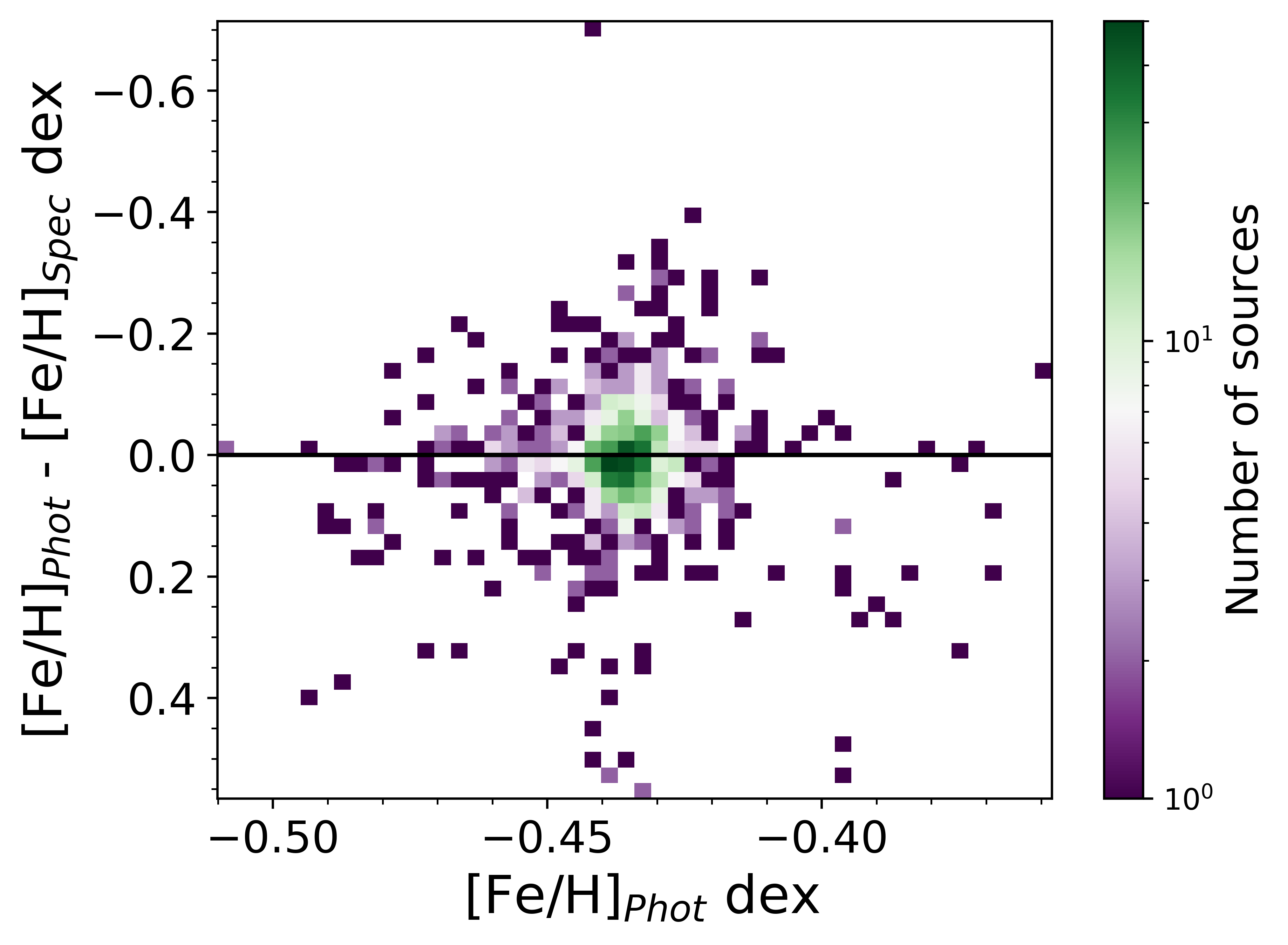}
	\caption{Same as Figure \ref{fig:rgb_recalib_hess} but for the supergiant sources.}
	\label{fig:bl_recalib_hess}
\end{figure}

Figure \ref{fig:rgb_hist_recalib} shows the metallicity distribution of the RGB sources in common between our sample and APOGEE. We fitted Gaussians to both histograms and estimated the peak and dispersion of each sample. The photometric metallicity peaks at --1.32$\pm$0.01 dex and the spectroscopic metallicity peaks at --0.90$\pm$0.02 dex, hence the peak difference is 0.43$\pm$0.02 dex. We added this difference to our estimated photometric metallicities. The bottom panel shows that after applying a shift both samples match. We also provide the metallicity distribution plots for the supergiant sources in the top panel of Figure \ref{fig:bl_hist_recalib}. The photometric and spectroscopic peak of the supergiants are --0.79$\pm$0.01 dex and --0.435$\pm$0.002 dex. The bottom panel shows that after adding 0.35$\pm$0.01 dex to the photometric metallicity, both samples match.\\
\indent To provide a quantitative idea of the residual bias and dispersion with respect to spectroscopic measures the top panel of Figure \ref{fig:rgb_recalib_hess} shows the density histograms of photometric metallicities vs spectroscopic metallicities of RGB stars in the Clouds. In the bottom panel, we plot the difference in the photometric and spectroscopic sample over the recalibrated metallicities. We fitted a line and the estimated slope is 0.45 with a constant of --0.33. We tested using these values to recalibrate our metallicities rather than just adding the peak difference estimates. Our overall gradients for the RGB sources of both Clouds did not change even by employing this recalibration, but we noticed larger uncertainties which would then propagate into the analysis of the gradient. The top panel of Figure \ref{fig:bl_recalib_hess} shows the photometric metallicities vs spectroscopic metallicities of the supergiants in the Clouds. In the bottom panel the difference is plotted against the recalibrated values. Here we do not see a clear slope for the RGB stars. The estimated slope for the supergiants is 0.13 with a constant value of --0.41. With this calibration, uncertainties for the supergiants would be ever larger.

\section{Estimation of peak difference based on radius}
\label{amplitudeshift}
\begin{figure*}
	\centering 
	\includegraphics[scale=0.35]{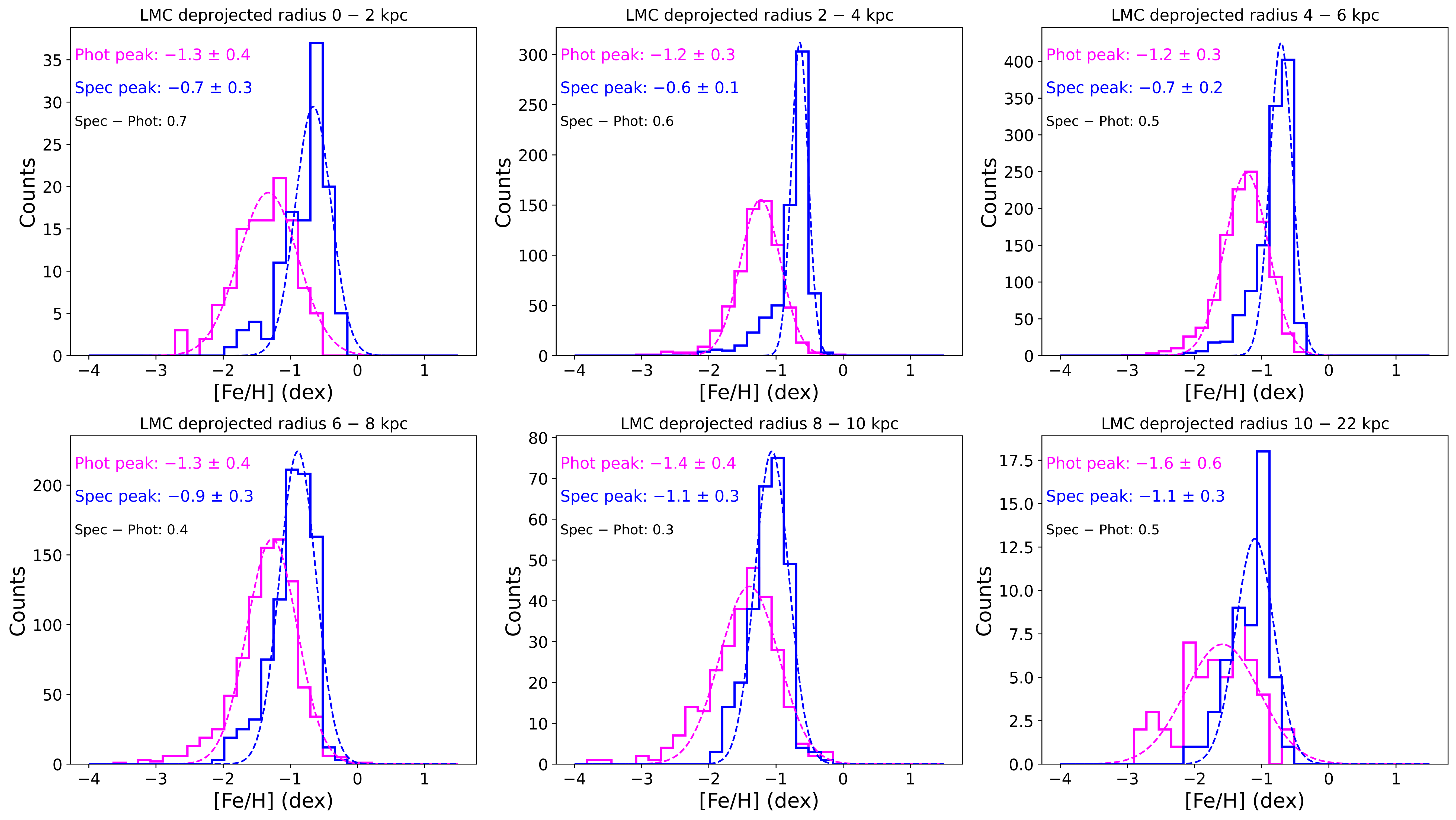}
	\includegraphics[scale=0.35]{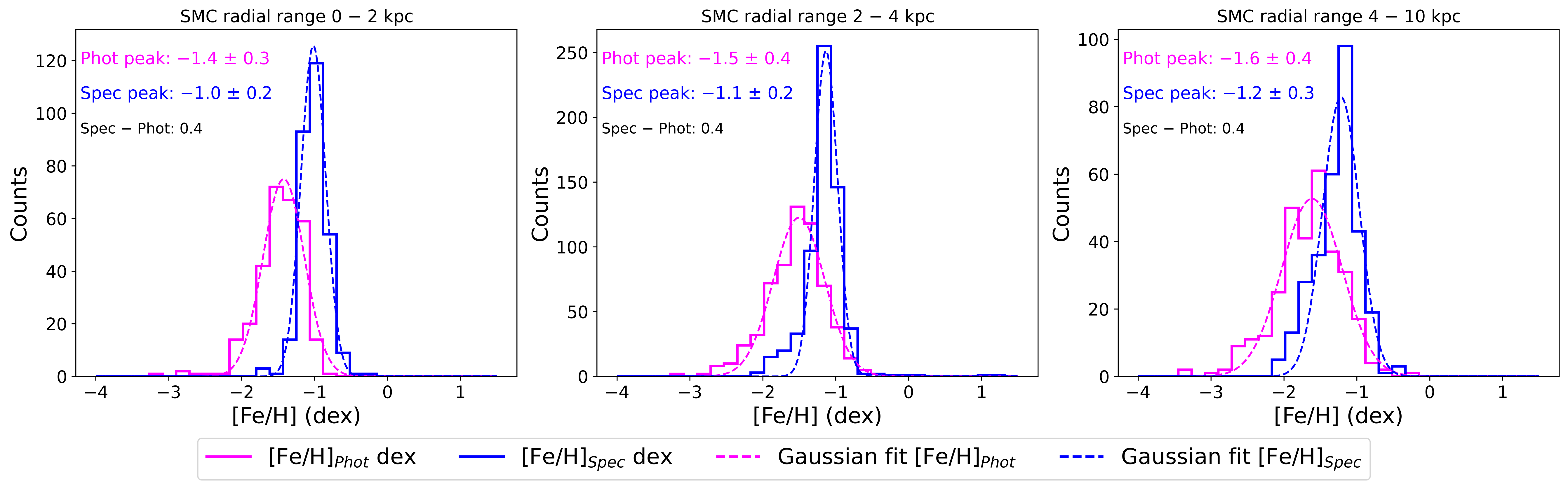}
	\includegraphics[scale=0.3]{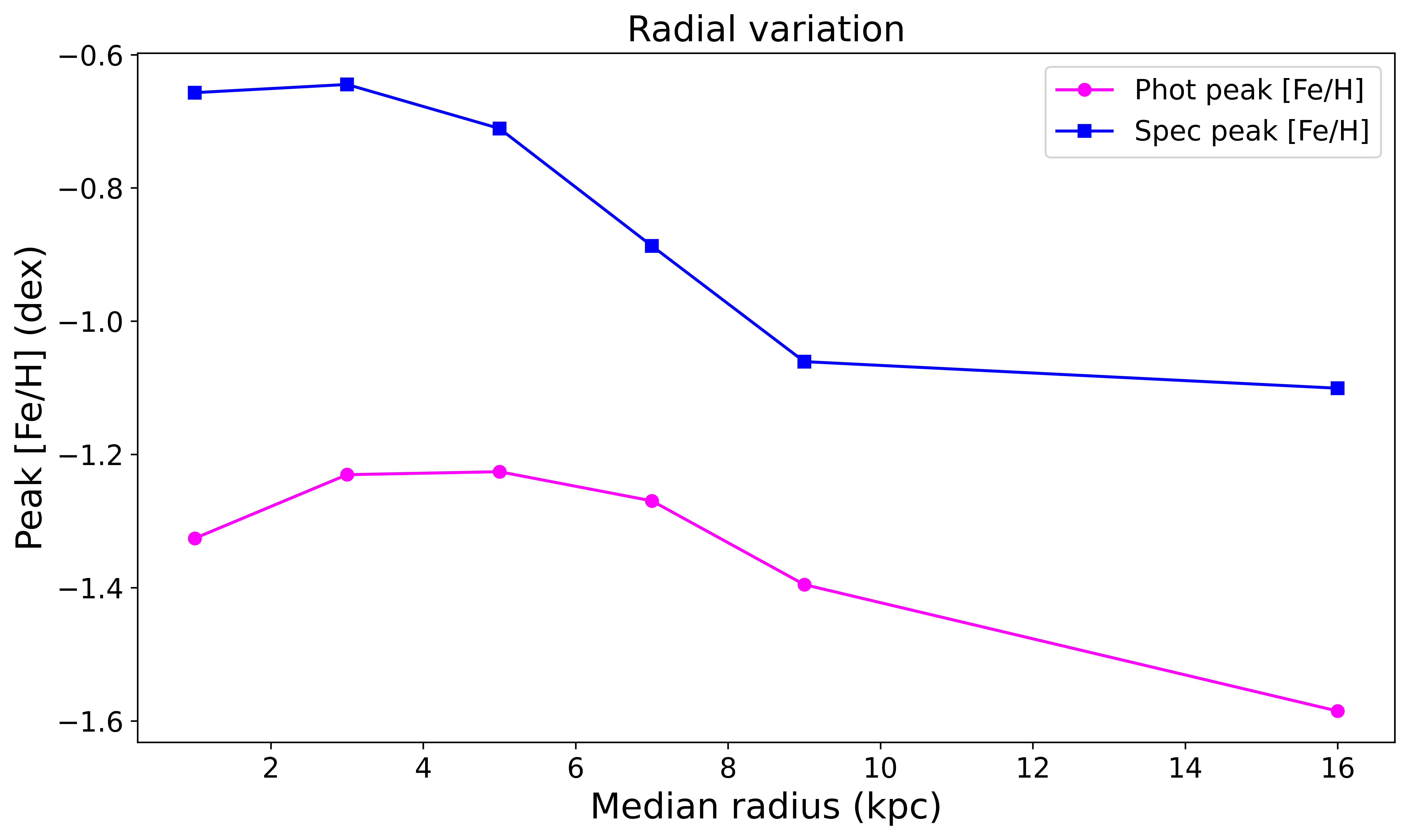}
	\includegraphics[scale=0.3]{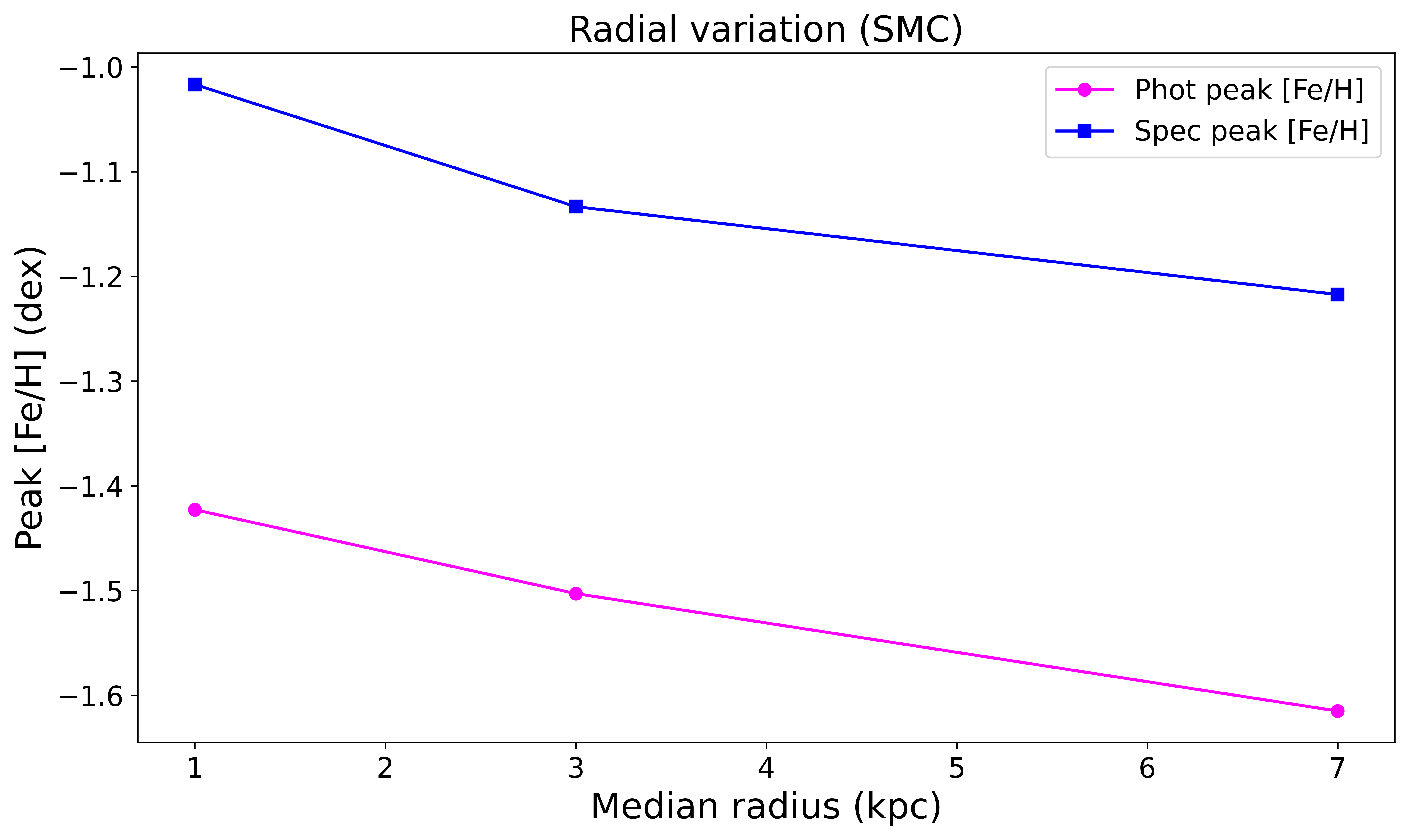}
	\caption{The top and middle rows display the peak differences between the photometric (Phot; blue) and spectroscopic (Spec; pink) RGB samples, calculated for various radial ranges from the centre of the LMC. The third row presents similar data for the RGB stars in the SMC. Both peaks are fitted, and the peak values are shown in the last row: on the left for the LMC and on the right for the SMC, which illustrates the radial variation of the shifts from the respective centres.}
	\label{fig:amplitude_shift_rgb}
\end{figure*}

\begin{figure*}
	\centering 
	\includegraphics[scale=0.35]{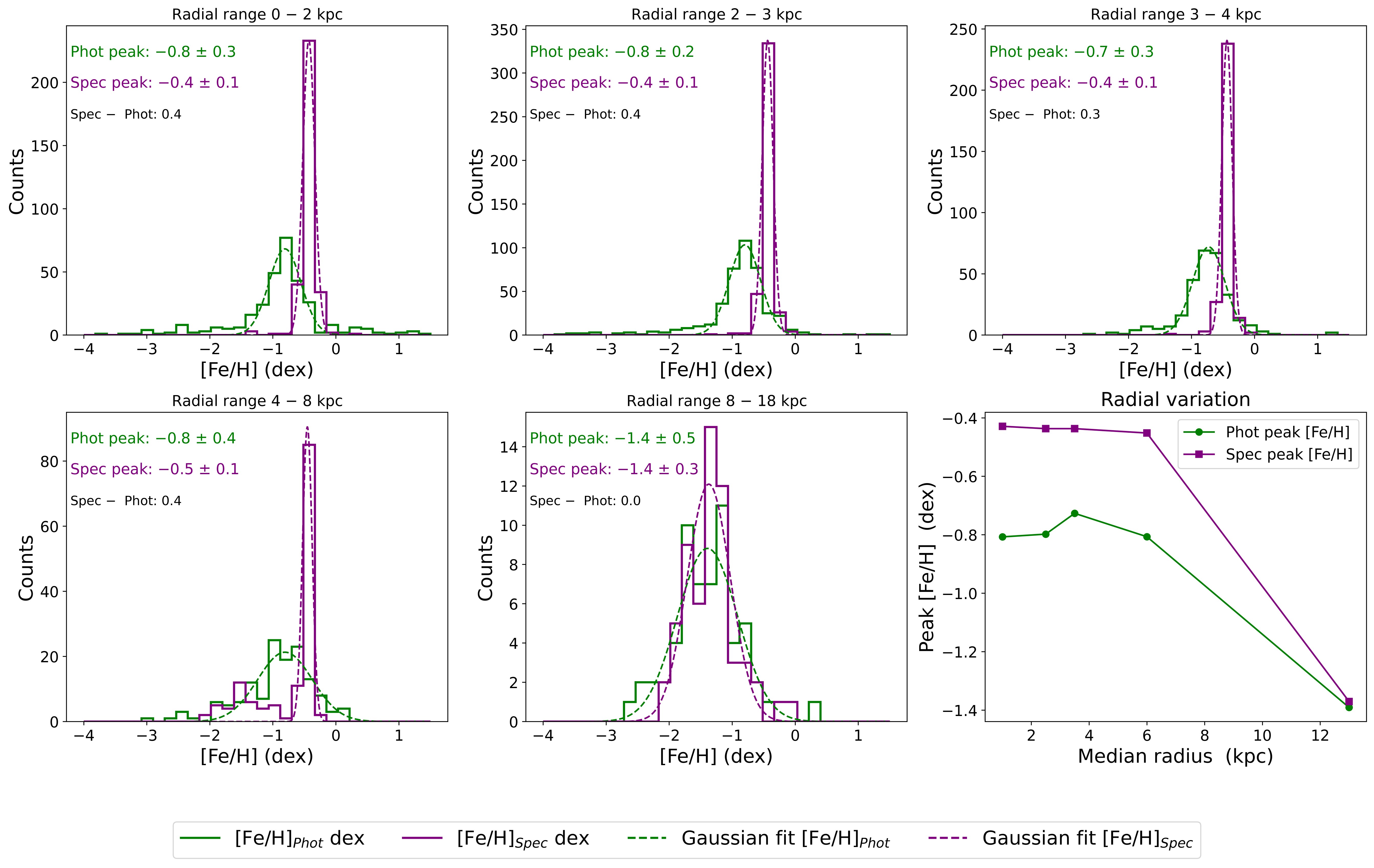}
	\caption{Same as Figure \ref{fig:amplitude_shift_rgb} but for the LMC and SMC supergiant stars together.}
	\label{fig:bl_amplitude_shift}
\end{figure*}

Crowding can cause blending and contamination of XP spectra, affecting our estimation of photometric metallicities, especially at the centre of the Clouds. To investigate whether we are observing any amplitude shift from the centre to the outskirts of the Clouds, we divided the area into various annular sub-regions from the respective centres. We plotted the distribution of photometric and spectroscopic metallicities for the LMC RGB stars in the top two rows of Figure \ref{fig:amplitude_shift_rgb}, and for the SMC RGB stars in the third row. The radial variation of the shifts from the respective centres are shown in the bottom-left for the LMC RGB stars and bottom-right for the SMC RGB stars. The same analysis was conducted for the supergiant stars and shown in Figure \ref{fig:bl_amplitude_shift}. For both distributions, we fitted Gaussians to identify the peak and standard deviation values for each radial region.

\section{Voronoi binning}
\label{voronoi}

\begin{figure*}[]
	\centering
	\includegraphics[width=0.8\columnwidth]{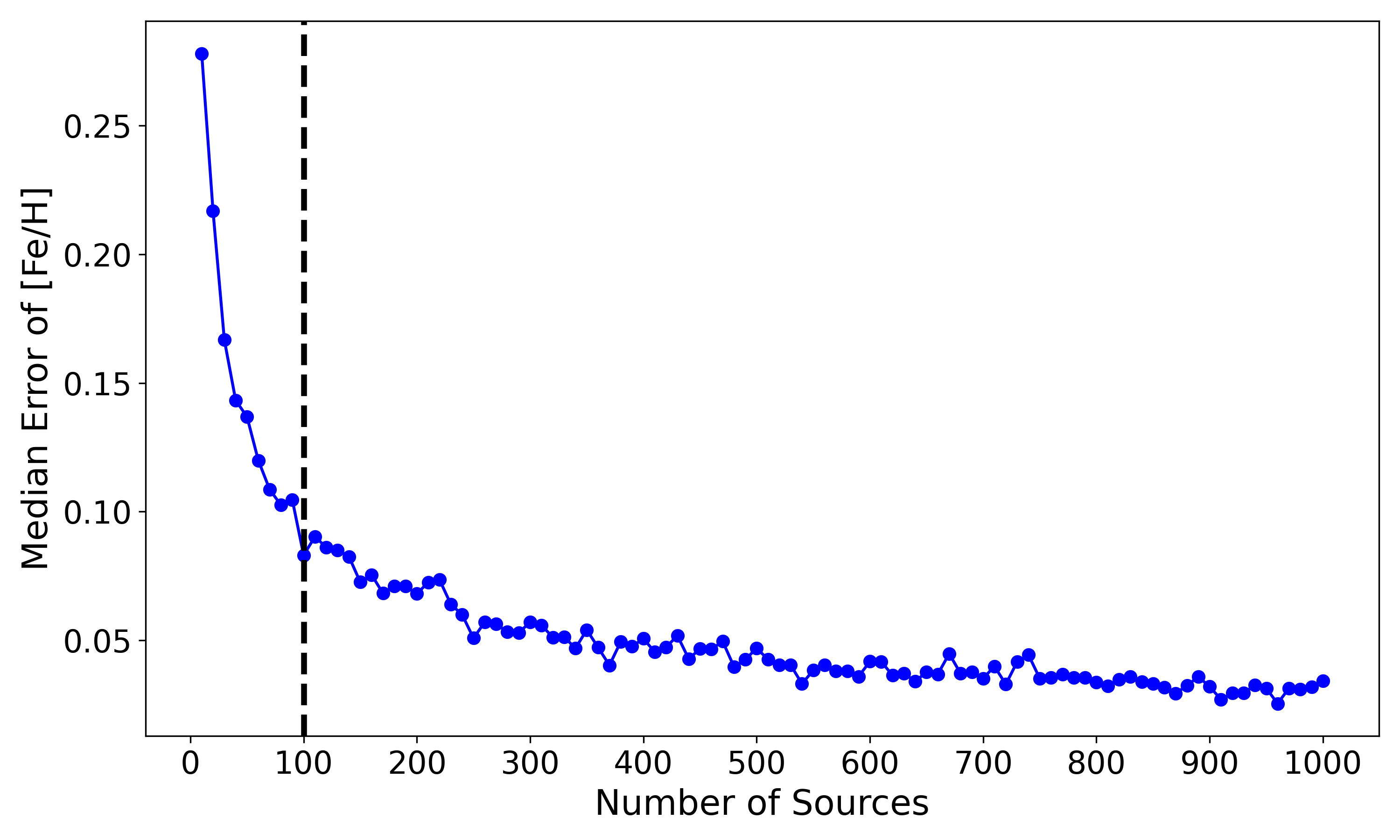}
	\includegraphics[width=0.8\columnwidth]{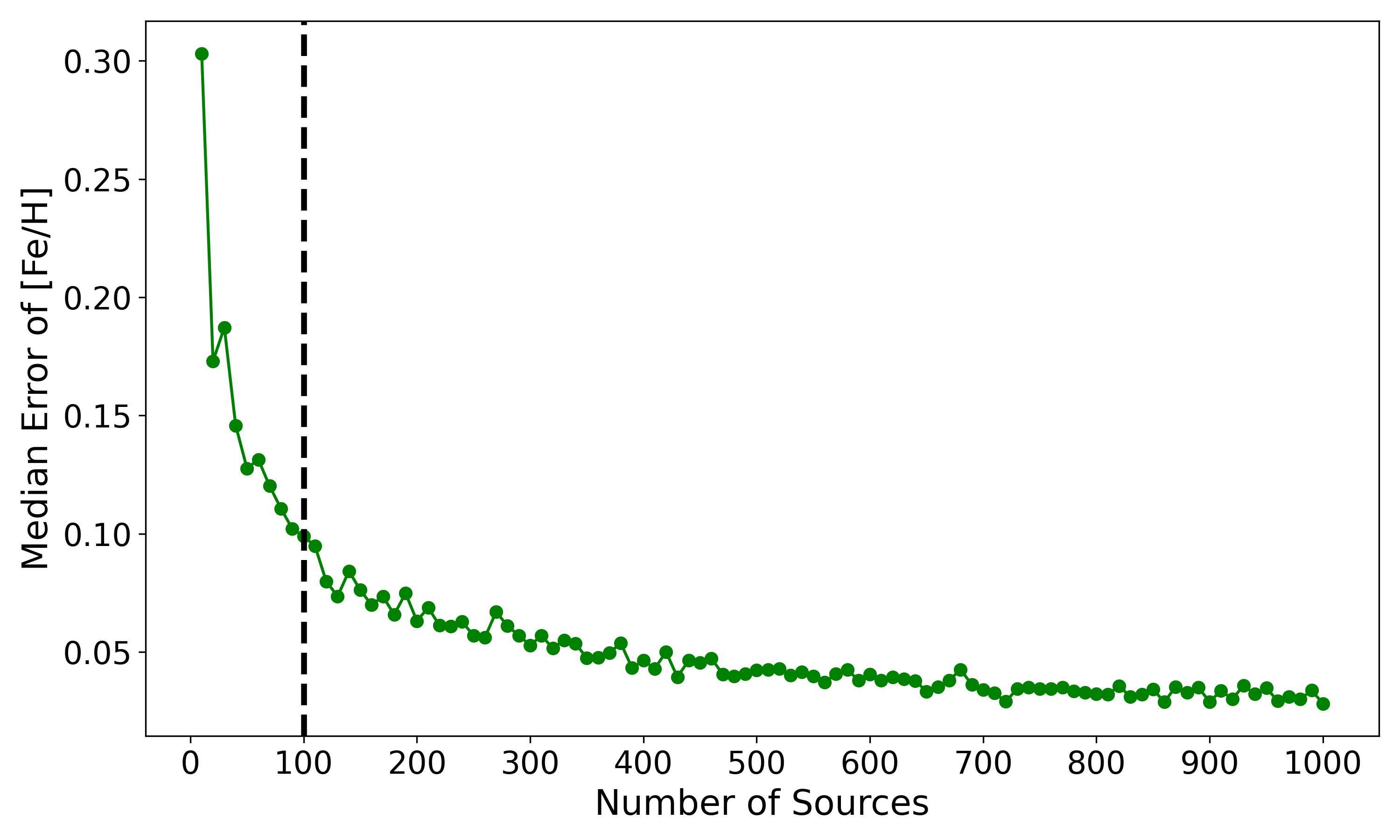}  
	\caption{The median error of [Fe/H] is plotted against the number of sources in each bin for RGB (left) and supergiant (right) stars in the Clouds. Dotted vertical lines at 100 sources indicate the limit adopted in our study.}
	\label{fig:mcmc}
\end{figure*}

\begin{figure*}
	\centering
	\vspace*{-0.4cm}
	\includegraphics[width=0.9\columnwidth]{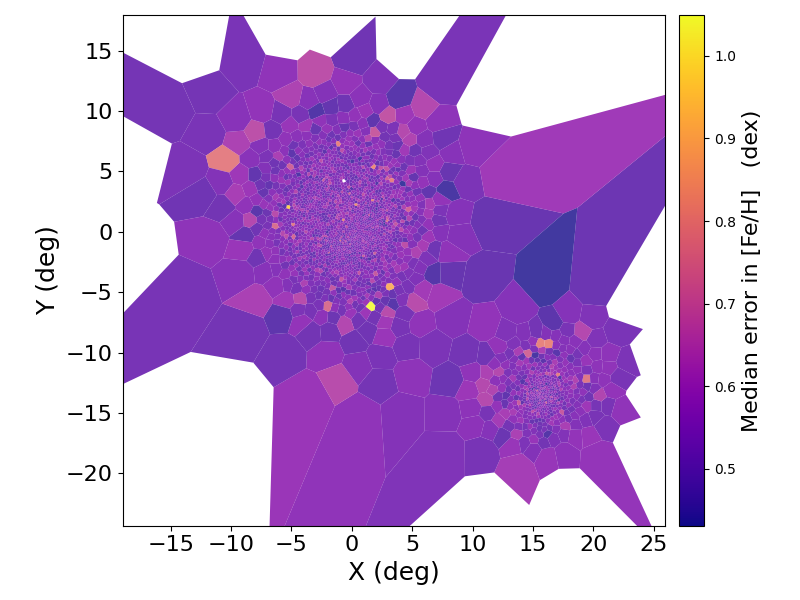}	
	\includegraphics[width=0.9\columnwidth]{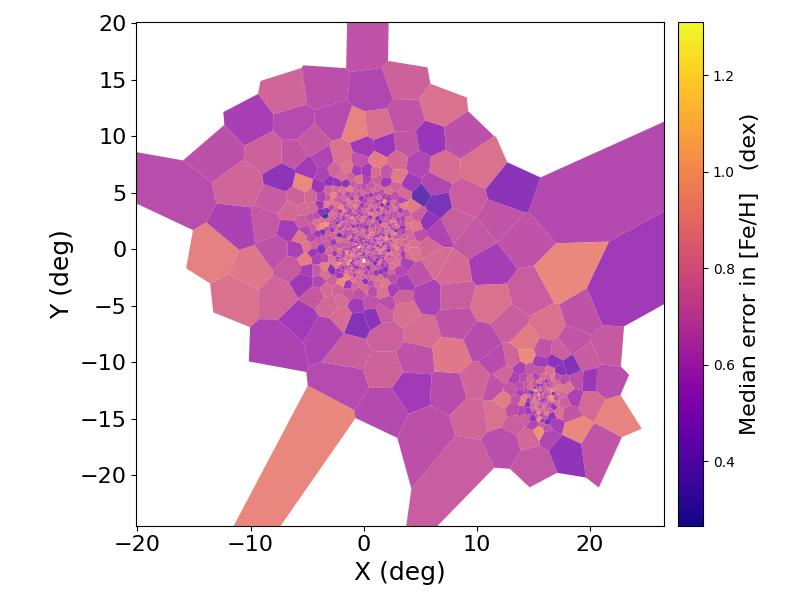}\\
	\caption{Maps showing the median error in the metallicity of RGB stars (left) and supergiants (right) of the Clouds. These maps are produced using the Voronoi binning method, where each bin has 100 sources. The colour-coding from purple to yellow indicates an increase in the median error.
	}
	\label{fig:voronoi_sigma_feh}
\end{figure*}

In this section, we outline the methodology used to determine an optimal number of sources for plotting the Voronoi metallicity maps of RGB stars and supergiants. To determine a sensible number of sources in each bin, we utilised MCMC simulations. The procedure was as follows: we drew random samples ranging from 10 to 1000, in steps of 10 from our entire dataset and conducted 100 simulations for each sample to determine the median error of [Fe/H]. The results produced a Gaussian distribution, which we fitted to estimate the peak and the associated dispersion, which is our median error for each run. The number of sources is plotted against this error for RGB stars and supergiants (see Figure \ref{fig:mcmc}). Our analysis revealed that for both RGB stars and supergiants, the error of the sample approached $\sim$0.1 dex when each bin included 100 sources. Consequently, we created a Voronoi metallicity map, ensuring that each of our bins contained 100 sources. Given the significant differences between our LMC and SMC samples, we generated the Voronoi metallicity maps separately for improved visualization. To facilitate easier interpretation of uncertainties associated with each bin of our Voronoi metallicity maps, we also produced a similar Voronoi binning map where we show the median of the [Fe/H] error in each of the bins. In Figure \ref{fig:voronoi_sigma_feh}, the top (RGB stars) and bottom (supergiants) panels show the median error of our metallicity distributions of the LMC (left) and the SMC (right). 

\section{Radial [Fe/H] gradient of the SMC and LMC sources in degrees}
\label{gradientdegree}
\begin{figure*}
    \centering
    \includegraphics[width=1\columnwidth]{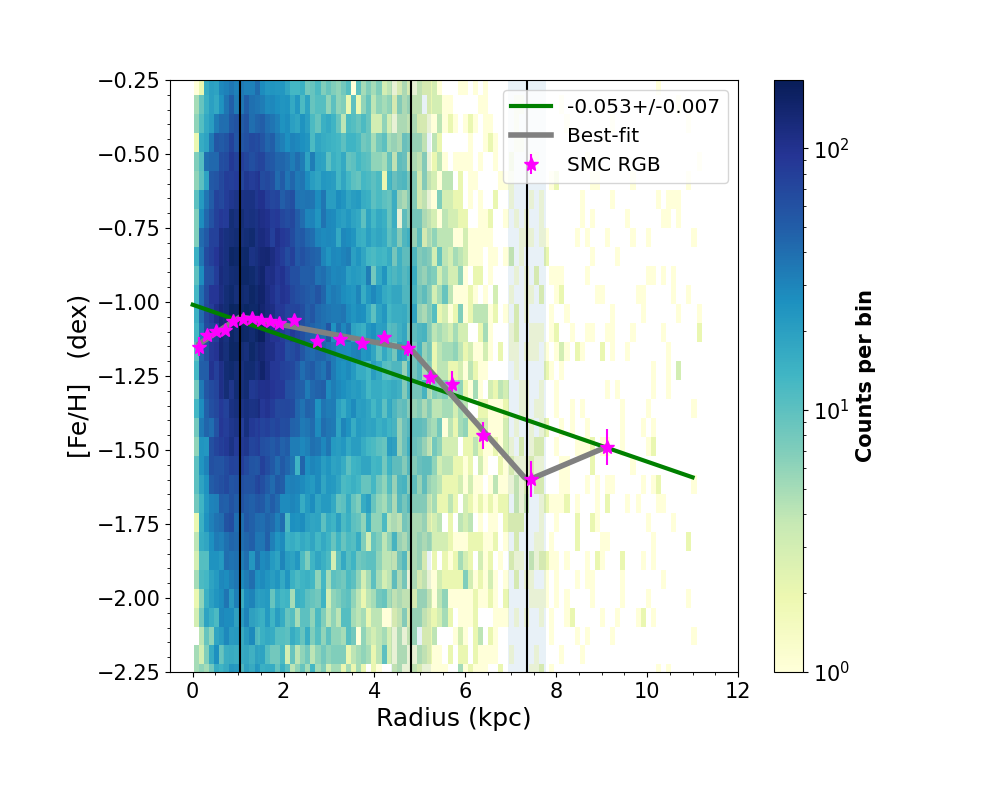}
    \includegraphics[width=1\columnwidth]{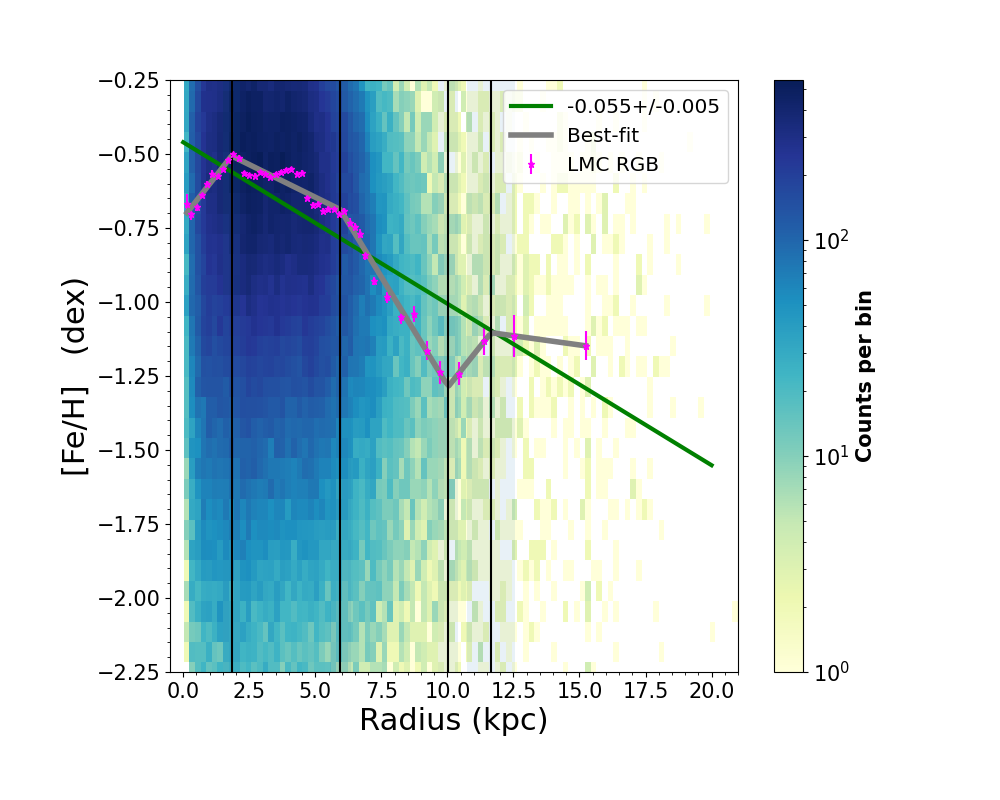} \\  
    \includegraphics[width=1\columnwidth]{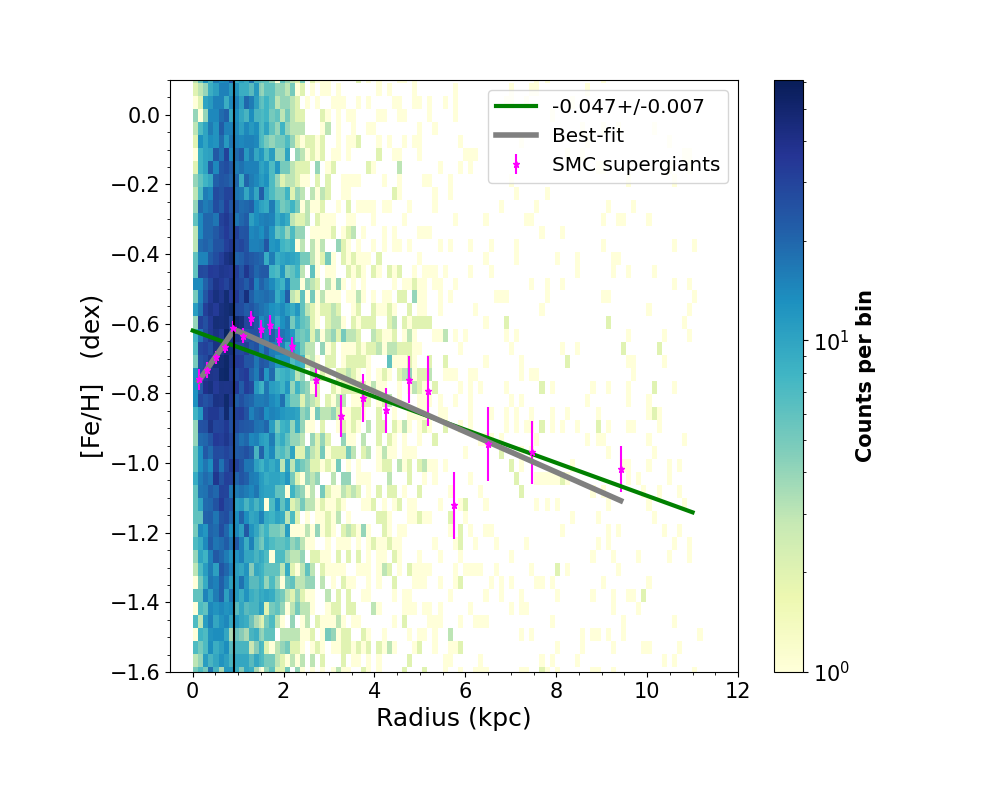}	
	\includegraphics[width=1\columnwidth]{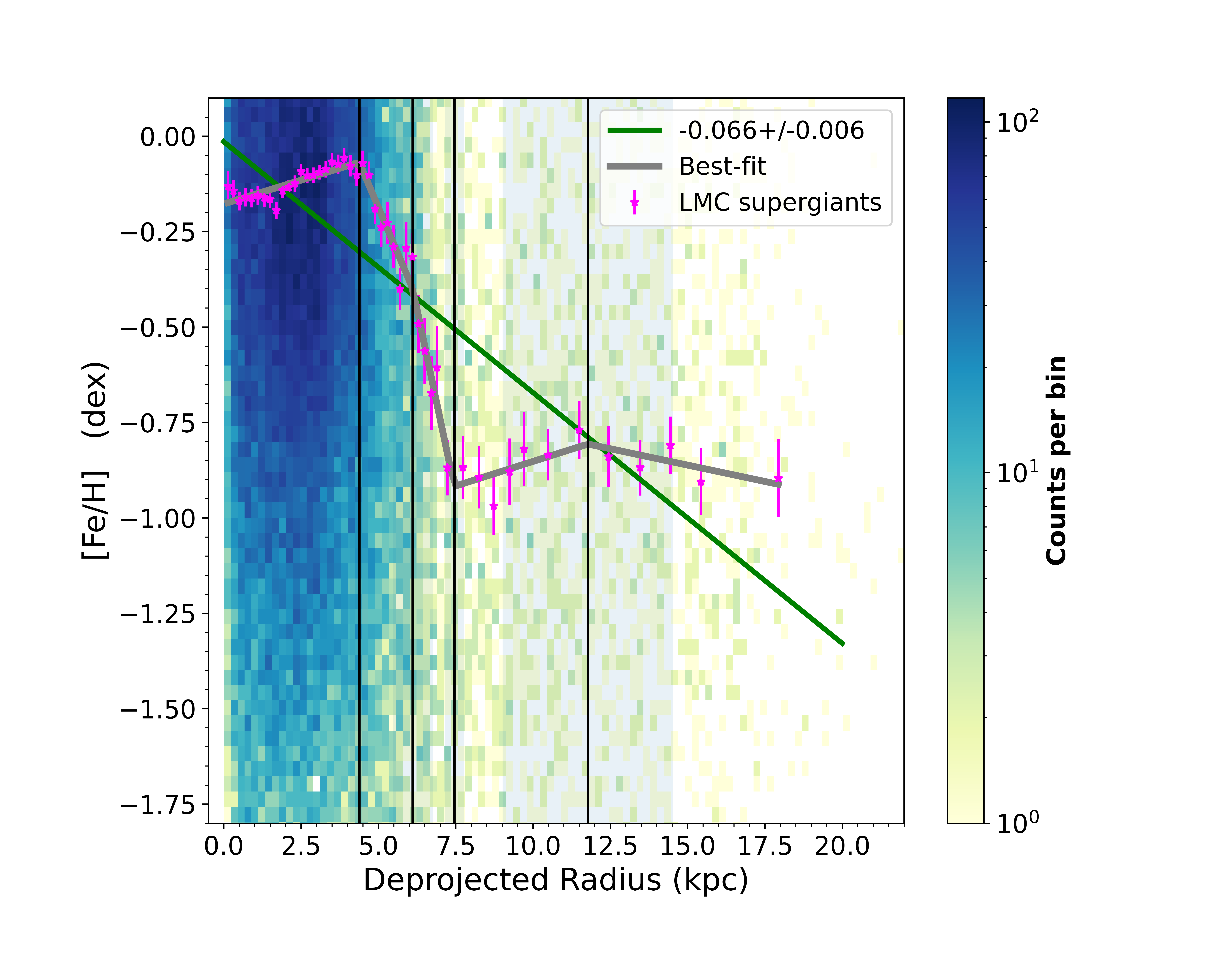}
  
     \caption{The radial metallicity gradient (in degrees) for the RGB (top) and supergiant (bottom) stars for SMC (left) and LMC (right). In all panels, the increasing stellar density is colour-coded from purple to yellow. The median metallicities of the annular regions are marked with pink stars along with their standard errors. Linear best-fits and segmented best-fits are shown in green and grey, respectively.  The black vertical lines show the division of different segments at the location of the breaks in the gradients. Underlying Hess diagrams show the source-density distributions.}
    \label{fig:feh_gradient}
\end{figure*}

The overall metallicity gradient as a function of distance (in degrees) is shown in Figure \ref{fig:feh_gradient}. For both populations in the SMC, we first radially bin the sources into 0.2 deg steps up to 2 deg from the centre. Then we used 0.5 deg steps up to 6 deg, 1 deg steps up to 8 deg and combine the rest of the sources until 11 deg into a single sub-region. The estimated overall metallicity gradient for the SMC RGB stars is --0.053$\pm$0.007 dex deg$^{-1}$ and the overall metallicity gradient estimated for the SMC supergiants is very similar, --0.047$\pm$0.008 dex deg$^{-1}$. For the LMC RGB stars and supergiants, we divide the sources into radial bins of 0.2 deg up to 7 deg from the centre and then use 0.5 deg bins up to 10 deg. Due to the lower number density, we made some adjustments for the RGB stars and supergiants. For the RGB stars, we use 1 deg bins up to 13 deg, while for the supergiants, we use 1 deg bins up to 16 deg. Finally, we combine the remaining sources into a single sub-region up to 20 deg. The estimated overall metallicity gradient for the LMC RGB stars is --0.055$\pm$0.005 dex deg$^{-1}$. The overall metallicity gradient estimated for the LMC supergiants is --0.066$\pm$0.006 dex deg$^{-1}$.

\begin{table}[]
\centering
\caption{Best-fit metallicity gradient of the SMC sources in degrees.}
\label{tab:smc_deg}
\begin{tabular}{cr|cr}
\hline \hline
\multicolumn{2}{c}{RGB}  & \multicolumn{2}{c}{Supergiants} \\ 
Distance & Gradient & Distance & Gradient \\
(deg) & (dex deg$^{-1}$) & (deg) & (dex deg$^{-1}$) \\
\hline
0 -- 1           & 0.100$\pm$0.034     & 0 -- 1           & 0.190$\pm$0.126             \\
1 -- 5           & --0.029$\pm$0.005  & R \textgreater 1 & 0.058$\pm$0.008\\
5 -- 7           & --0.174$\pm$0.024  &                  &                 \\
R \textgreater 7 & 0.065$\pm$0.017   &                  &                 \\
0 -- 11          & --0.053$\pm$0.007 & 0 -- 11          & --0.047$\pm$0.007       \\     
\hline
\end{tabular}
\end{table}

\begin{table}[]
\centering
\caption{Best-fit metallicity gradient of the LMC sources in degrees.}
\label{tab:lmc_deg}
\begin{tabular}{cr|cr}
\hline \hline
\multicolumn{2}{c}{RGB}                           & \multicolumn{2}{c}{Supergiants}                   \\
 Distance & Gradient & Distance & Gradient \\
(deg) & (dex deg$^{-1}$) & (deg) & (dex deg$^{-1}$) \\
\hline
0 -- 2            & 0.114$\pm$0.02    & 0 -- 5            & 0.025$\pm$0.007   \\
2 -- 5.5          & --0.044$\pm$0.005  & 5 -- 6            & --0.192$\pm$0.022  \\
5.5 -- 10         & --0.147$\pm$0.008  & 6 -- 8            & --0.385$\pm$0.076  \\
10 -- 12          & 0.112$\pm$0.043   & 8 -- 12           & 0.026$\pm$0.018   \\
R \textgreater 12 & --0.012$\pm$0.016  & R \textgreater 12 & --0.017$\pm$0.008  \\
0 -- 20           & --0.055$\pm$0.005 & 0 -- 20           & --0.066$\pm$0.006 \\
\hline
\end{tabular}
\end{table}

\end{appendix}
\end{document}